\documentclass[conference]{IEEEtran}

\def\review{0} 

\usepackage{tikz}  
\usetikzlibrary{spy}
\usepackage{color, colortbl}
\usepackage{pgfplots}
\pgfplotsset{compat=newest}
\usepackage{multicol}
\usepackage{circuitikz}
\usepackage[nolist]{acronym}
\usepgfplotslibrary{groupplots}
\usetikzlibrary{shapes}
\usepackage{siunitx}  
\usepackage{graphicx}
\usepgfplotslibrary{patchplots}
\usepackage{amsmath,amssymb}
\usepackage{subfig}
 \usepackage{float}
\usepackage{cite}
\usepackage{amsmath}
\usetikzlibrary{angles, quotes,calc,patterns}
\usepgfplotslibrary{fillbetween}

\usepackage{ifthen}
\usepackage{algorithm}
\usepackage[noend]{algpseudocode}
\usetikzlibrary{matrix,calc}
\usepackage{trfsigns}

\usepackage{textcomp} 
\usepackage{nicefrac}
\usepackage{multirow}
\usepackage{booktabs}
\usepackage[shortcuts,acronym]{glossaries}

\if\review1
\usepackage{todonotes}
\else
\usepackage[disable]{todonotes}
\fi

\renewcommand{\vec}[1]{\mathbf{#1}}

\definecolor{uni_apfelgruen}{cmyk}{.5, 0, 1, 0}
\definecolor{uni_mittelblau}{cmyk}{1, 0.4, 0, 0}
\definecolor{uni_bulletblau}{RGB}{49,99,183}
\definecolor{uni_gelb}{cmyk}{0, 0.1, 1, 0}
\definecolor{uni_rot}{cmyk}{0, 1, 1, 0}
\definecolor{uniblauHell}{RGB}{0,190,255}
\definecolor{uniblauDunkel}{RGB}{0,65,145}
\definecolor{unigrau}{RGB}{51,51,51}
\definecolor{darkgray176}{RGB}{176,176,176}
\definecolor{lavenderplot}{RGB}{191,148,228}
\definecolor{coralplot}{RGB}{255,127,80}
\definecolor{cyanplot}{RGB}{37,219,168}

\newacronym{sap}{SAP}{sensing access point}
\newacronym{semf}{SeMF}{sensing management function}
\newacronym{lmf}{LMF}{location management function}
\newacronym{adc}{ADC}{analog-to-digital converter}
\newacronym{snr}{SNR}{signal-to-noise ratio}
\newacronym{rcs}{RCS}{radar cross-section}
\newacronym{nf}{NF}{noise figure}
\newacronym{tx}{Tx}{transmitter}
\newacronym{rx}{Rx}{receiver}
\newacronym{awgn}{AWGN}{additive white Gaussian noise}
\newacronym{cir}{CIR}{channel impulse response}
\newacronym{ctf}{CTF}{channel transfer function}
\newacronym{dft}{DFT}{discrete Fourier transform}
\newacronym{cfar}{CFAR}{constant false alarm rate}
\newacronym{ofdm}{OFDM}{orthogonal frequency-division multiplexing}
\newacronym{mimo}{MIMO}{multiple-input and multiple-output}
\newacronym{ula}{ULA}{uniform linear array}
\newacronym{isac}{ISAC}{integrated sensing and communication}
\newacronym{dl}{DL}{deep learning}

\title{Multi-Target Localization in Multi-Static Integrated Sensing and Communication Deployments}

\author{
\IEEEauthorblockN{Maximilian Bauhofer\IEEEauthorrefmark{1}, Silvio Mandelli\IEEEauthorrefmark{2}, 
Marcus Henninger\IEEEauthorrefmark{2}\IEEEauthorrefmark{1}, 
Thorsten Wild\IEEEauthorrefmark{2} and Stephan ten Brink\IEEEauthorrefmark{1}}
\IEEEauthorblockA{
\IEEEauthorrefmark{1}Institute of Telecommunications, Pfaffenwaldring 47, University of  Stuttgart, 70659 Stuttgart, Germany 
\\\{bauhofer, tenbrink\}@inue.uni-stuttgart.de
}
\IEEEauthorblockA{\IEEEauthorrefmark{2}Nokia Bell Labs, 70469 Stuttgart, Germany}}

\newcommand\blfootnote[1]{%
  \begingroup
  \renewcommand\thefootnote{}\footnote{#1}%
  \addtocounter{footnote}{-1}%
  \endgroup
}

\begin{document}




\pgfplotsset{
    colormap={jet_inue}{
        rgb=(1, 1, 1)
        rgb=(0.99804, 0.99804, 0.99905)
        rgb=(0.99609, 0.99609, 0.99813)
        rgb=(0.99413, 0.99413, 0.99725)
        rgb=(0.99217, 0.99217, 0.99639)
        rgb=(0.99022, 0.99022, 0.99557)
        rgb=(0.98826, 0.98826, 0.99477)
        rgb=(0.9863, 0.9863, 0.99401)
        rgb=(0.98434, 0.98434, 0.99327)
        rgb=(0.98239, 0.98239, 0.99257)
        rgb=(0.98043, 0.98043, 0.9919)
        rgb=(0.97847, 0.97847, 0.99125)
        rgb=(0.97652, 0.97652, 0.99064)
        rgb=(0.97456, 0.97456, 0.99006)
        rgb=(0.9726, 0.9726, 0.98951)
        rgb=(0.97065, 0.97065, 0.98899)
        rgb=(0.96869, 0.96869, 0.9885)
        rgb=(0.96673, 0.96673, 0.98804)
        rgb=(0.96477, 0.96477, 0.98762)
        rgb=(0.96282, 0.96282, 0.98722)
        rgb=(0.96086, 0.96086, 0.98685)
        rgb=(0.9589, 0.9589, 0.98652)
        rgb=(0.95695, 0.95695, 0.98621)
        rgb=(0.95499, 0.95499, 0.98593)
        rgb=(0.95303, 0.95303, 0.98569)
        rgb=(0.95108, 0.95108, 0.98548)
        rgb=(0.94912, 0.94912, 0.98529)
        rgb=(0.94716, 0.94716, 0.98514)
        rgb=(0.94521, 0.94521, 0.98502)
        rgb=(0.94325, 0.94325, 0.98493)
        rgb=(0.94129, 0.94129, 0.98486)
        rgb=(0.93933, 0.93933, 0.98483)
        rgb=(0.93738, 0.93738, 0.98483)
        rgb=(0.93542, 0.93542, 0.98486)
        rgb=(0.93346, 0.93346, 0.98493)
        rgb=(0.93151, 0.93151, 0.98502)
        rgb=(0.92955, 0.92955, 0.98514)
        rgb=(0.92759, 0.92759, 0.98529)
        rgb=(0.92564, 0.92564, 0.98548)
        rgb=(0.92368, 0.92368, 0.98569)
        rgb=(0.92172, 0.92172, 0.98593)
        rgb=(0.91977, 0.91977, 0.98621)
        rgb=(0.91781, 0.91781, 0.98652)
        rgb=(0.91585, 0.91585, 0.98685)
        rgb=(0.91389, 0.91389, 0.98722)
        rgb=(0.91194, 0.91194, 0.98762)
        rgb=(0.90998, 0.90998, 0.98804)
        rgb=(0.90802, 0.90802, 0.9885)
        rgb=(0.90607, 0.90607, 0.98899)
        rgb=(0.90411, 0.90411, 0.98951)
        rgb=(0.90215, 0.90215, 0.99006)
        rgb=(0.9002, 0.9002, 0.99064)
        rgb=(0.89824, 0.89824, 0.99125)
        rgb=(0.89628, 0.89628, 0.9919)
        rgb=(0.89432, 0.89432, 0.99257)
        rgb=(0.89237, 0.89237, 0.99327)
        rgb=(0.89041, 0.89041, 0.99401)
        rgb=(0.88845, 0.88845, 0.99477)
        rgb=(0.8865, 0.8865, 0.99557)
        rgb=(0.88454, 0.88454, 0.99639)
        rgb=(0.88258, 0.88258, 0.99725)
        rgb=(0.88063, 0.88063, 0.99813)
        rgb=(0.87867, 0.87867, 0.99905)
        rgb=(0.87671, 0.87671, 1)
        rgb=(0.87476, 0.87573, 1)
        rgb=(0.8728, 0.87479, 1)
        rgb=(0.87084, 0.87387, 1)
        rgb=(0.86888, 0.87298, 1)
        rgb=(0.86693, 0.87213, 1)
        rgb=(0.86497, 0.8713, 1)
        rgb=(0.86301, 0.87051, 1)
        rgb=(0.86106, 0.86974, 1)
        rgb=(0.8591, 0.86901, 1)
        rgb=(0.85714, 0.8683, 1)
        rgb=(0.85519, 0.86763, 1)
        rgb=(0.85323, 0.86699, 1)
        rgb=(0.85127, 0.86638, 1)
        rgb=(0.84932, 0.8658, 1)
        rgb=(0.84736, 0.86525, 1)
        rgb=(0.8454, 0.86473, 1)
        rgb=(0.84344, 0.86424, 1)
        rgb=(0.84149, 0.86378, 1)
        rgb=(0.83953, 0.86335, 1)
        rgb=(0.83757, 0.86295, 1)
        rgb=(0.83562, 0.86259, 1)
        rgb=(0.83366, 0.86225, 1)
        rgb=(0.8317, 0.86194, 1)
        rgb=(0.82975, 0.86167, 1)
        rgb=(0.82779, 0.86142, 1)
        rgb=(0.82583, 0.86121, 1)
        rgb=(0.82387, 0.86103, 1)
        rgb=(0.82192, 0.86087, 1)
        rgb=(0.81996, 0.86075, 1)
        rgb=(0.818, 0.86066, 1)
        rgb=(0.81605, 0.8606, 1)
        rgb=(0.81409, 0.86057, 1)
        rgb=(0.81213, 0.86057, 1)
        rgb=(0.81018, 0.8606, 1)
        rgb=(0.80822, 0.86066, 1)
        rgb=(0.80626, 0.86075, 1)
        rgb=(0.80431, 0.86087, 1)
        rgb=(0.80235, 0.86103, 1)
        rgb=(0.80039, 0.86121, 1)
        rgb=(0.79843, 0.86142, 1)
        rgb=(0.79648, 0.86167, 1)
        rgb=(0.79452, 0.86194, 1)
        rgb=(0.79256, 0.86225, 1)
        rgb=(0.79061, 0.86259, 1)
        rgb=(0.78865, 0.86295, 1)
        rgb=(0.78669, 0.86335, 1)
        rgb=(0.78474, 0.86378, 1)
        rgb=(0.78278, 0.86424, 1)
        rgb=(0.78082, 0.86473, 1)
        rgb=(0.77886, 0.86525, 1)
        rgb=(0.77691, 0.8658, 1)
        rgb=(0.77495, 0.86638, 1)
        rgb=(0.77299, 0.86699, 1)
        rgb=(0.77104, 0.86763, 1)
        rgb=(0.76908, 0.8683, 1)
        rgb=(0.76712, 0.86901, 1)
        rgb=(0.76517, 0.86974, 1)
        rgb=(0.76321, 0.87051, 1)
        rgb=(0.76125, 0.8713, 1)
        rgb=(0.7593, 0.87213, 1)
        rgb=(0.75734, 0.87298, 1)
        rgb=(0.75538, 0.87387, 1)
        rgb=(0.75342, 0.87479, 1)
        rgb=(0.75147, 0.87573, 1)
        rgb=(0.74951, 0.87671, 1)
        rgb=(0.74755, 0.87772, 1)
        rgb=(0.7456, 0.87876, 1)
        rgb=(0.74364, 0.87983, 1)
        rgb=(0.74168, 0.88093, 1)
        rgb=(0.73973, 0.88206, 1)
        rgb=(0.73777, 0.88323, 1)
        rgb=(0.73581, 0.88442, 1)
        rgb=(0.73386, 0.88564, 1)
        rgb=(0.7319, 0.88689, 1)
        rgb=(0.72994, 0.88818, 1)
        rgb=(0.72798, 0.88949, 1)
        rgb=(0.72603, 0.89084, 1)
        rgb=(0.72407, 0.89222, 1)
        rgb=(0.72211, 0.89362, 1)
        rgb=(0.72016, 0.89506, 1)
        rgb=(0.7182, 0.89653, 1)
        rgb=(0.71624, 0.89802, 1)
        rgb=(0.71429, 0.89955, 1)
        rgb=(0.71233, 0.90111, 1)
        rgb=(0.71037, 0.9027, 1)
        rgb=(0.70841, 0.90432, 1)
        rgb=(0.70646, 0.90597, 1)
        rgb=(0.7045, 0.90766, 1)
        rgb=(0.70254, 0.90937, 1)
        rgb=(0.70059, 0.91111, 1)
        rgb=(0.69863, 0.91289, 1)
        rgb=(0.69667, 0.91469, 1)
        rgb=(0.69472, 0.91652, 1)
        rgb=(0.69276, 0.91839, 1)
        rgb=(0.6908, 0.92028, 1)
        rgb=(0.68885, 0.92221, 1)
        rgb=(0.68689, 0.92417, 1)
        rgb=(0.68493, 0.92616, 1)
        rgb=(0.68297, 0.92817, 1)
        rgb=(0.68102, 0.93022, 1)
        rgb=(0.67906, 0.9323, 1)
        rgb=(0.6771, 0.93441, 1)
        rgb=(0.67515, 0.93655, 1)
        rgb=(0.67319, 0.93872, 1)
        rgb=(0.67123, 0.94092, 1)
        rgb=(0.66928, 0.94316, 1)
        rgb=(0.66732, 0.94542, 1)
        rgb=(0.66536, 0.94771, 1)
        rgb=(0.66341, 0.95004, 1)
        rgb=(0.66145, 0.95239, 1)
        rgb=(0.65949, 0.95478, 1)
        rgb=(0.65753, 0.95719, 1)
        rgb=(0.65558, 0.95964, 1)
        rgb=(0.65362, 0.96211, 1)
        rgb=(0.65166, 0.96462, 1)
        rgb=(0.64971, 0.96716, 1)
        rgb=(0.64775, 0.96973, 1)
        rgb=(0.64579, 0.97233, 1)
        rgb=(0.64384, 0.97496, 1)
        rgb=(0.64188, 0.97762, 1)
        rgb=(0.63992, 0.98031, 1)
        rgb=(0.63796, 0.98303, 1)
        rgb=(0.63601, 0.98578, 1)
        rgb=(0.63405, 0.98856, 1)
        rgb=(0.63209, 0.99138, 1)
        rgb=(0.63014, 0.99422, 1)
        rgb=(0.62818, 0.9971, 1)
        rgb=(0.62622, 1, 1)
        rgb=(0.6272, 1, 0.99706)
        rgb=(0.62821, 1, 0.9941)
        rgb=(0.62925, 1, 0.9911)
        rgb=(0.63032, 1, 0.98807)
        rgb=(0.63142, 1, 0.98502)
        rgb=(0.63255, 1, 0.98193)
        rgb=(0.63371, 1, 0.97881)
        rgb=(0.63491, 1, 0.97566)
        rgb=(0.63613, 1, 0.97248)
        rgb=(0.63738, 1, 0.96927)
        rgb=(0.63867, 1, 0.96603)
        rgb=(0.63998, 1, 0.96276)
        rgb=(0.64133, 1, 0.95945)
        rgb=(0.6427, 1, 0.95612)
        rgb=(0.64411, 1, 0.95276)
        rgb=(0.64555, 1, 0.94936)
        rgb=(0.64702, 1, 0.94594)
        rgb=(0.64851, 1, 0.94248)
        rgb=(0.65004, 1, 0.939)
        rgb=(0.6516, 1, 0.93548)
        rgb=(0.65319, 1, 0.93193)
        rgb=(0.65481, 1, 0.92836)
        rgb=(0.65646, 1, 0.92475)
        rgb=(0.65815, 1, 0.92111)
        rgb=(0.65986, 1, 0.91744)
        rgb=(0.6616, 1, 0.91374)
        rgb=(0.66337, 1, 0.91001)
        rgb=(0.66518, 1, 0.90625)
        rgb=(0.66701, 1, 0.90246)
        rgb=(0.66888, 1, 0.89864)
        rgb=(0.67077, 1, 0.89478)
        rgb=(0.6727, 1, 0.8909)
        rgb=(0.67466, 1, 0.88699)
        rgb=(0.67665, 1, 0.88304)
        rgb=(0.67866, 1, 0.87907)
        rgb=(0.68071, 1, 0.87506)
        rgb=(0.68279, 1, 0.87102)
        rgb=(0.6849, 1, 0.86696)
        rgb=(0.68704, 1, 0.86286)
        rgb=(0.68921, 1, 0.85873)
        rgb=(0.69141, 1, 0.85457)
        rgb=(0.69365, 1, 0.85039)
        rgb=(0.69591, 1, 0.84617)
        rgb=(0.6982, 1, 0.84192)
        rgb=(0.70053, 1, 0.83763)
        rgb=(0.70288, 1, 0.83332)
        rgb=(0.70527, 1, 0.82898)
        rgb=(0.70768, 1, 0.82461)
        rgb=(0.71013, 1, 0.82021)
        rgb=(0.7126, 1, 0.81577)
        rgb=(0.71511, 1, 0.81131)
        rgb=(0.71765, 1, 0.80681)
        rgb=(0.72022, 1, 0.80229)
        rgb=(0.72282, 1, 0.79773)
        rgb=(0.72545, 1, 0.79314)
        rgb=(0.72811, 1, 0.78853)
        rgb=(0.7308, 1, 0.78388)
        rgb=(0.73352, 1, 0.7792)
        rgb=(0.73627, 1, 0.77449)
        rgb=(0.73905, 1, 0.76975)
        rgb=(0.74187, 1, 0.76498)
        rgb=(0.74471, 1, 0.76018)
        rgb=(0.74758, 1, 0.75535)
        rgb=(0.75049, 1, 0.75049)
        rgb=(0.75342, 1, 0.7456)
        rgb=(0.75639, 1, 0.74067)
        rgb=(0.75939, 1, 0.73572)
        rgb=(0.76241, 1, 0.73074)
        rgb=(0.76547, 1, 0.72572)
        rgb=(0.76856, 1, 0.72068)
        rgb=(0.77168, 1, 0.7156)
        rgb=(0.77483, 1, 0.71049)
        rgb=(0.77801, 1, 0.70536)
        rgb=(0.78122, 1, 0.70019)
        rgb=(0.78446, 1, 0.69499)
        rgb=(0.78773, 1, 0.68976)
        rgb=(0.79103, 1, 0.6845)
        rgb=(0.79437, 1, 0.67921)
        rgb=(0.79773, 1, 0.67389)
        rgb=(0.80113, 1, 0.66854)
        rgb=(0.80455, 1, 0.66316)
        rgb=(0.80801, 1, 0.65775)
        rgb=(0.81149, 1, 0.65231)
        rgb=(0.81501, 1, 0.64683)
        rgb=(0.81855, 1, 0.64133)
        rgb=(0.82213, 1, 0.63579)
        rgb=(0.82574, 1, 0.63023)
        rgb=(0.82938, 1, 0.62463)
        rgb=(0.83305, 1, 0.61901)
        rgb=(0.83675, 1, 0.61335)
        rgb=(0.84048, 1, 0.60766)
        rgb=(0.84424, 1, 0.60194)
        rgb=(0.84803, 1, 0.5962)
        rgb=(0.85185, 1, 0.59042)
        rgb=(0.85571, 1, 0.58461)
        rgb=(0.85959, 1, 0.57877)
        rgb=(0.8635, 1, 0.5729)
        rgb=(0.86745, 1, 0.56699)
        rgb=(0.87142, 1, 0.56106)
        rgb=(0.87543, 1, 0.5551)
        rgb=(0.87946, 1, 0.54911)
        rgb=(0.88353, 1, 0.54308)
        rgb=(0.88763, 1, 0.53703)
        rgb=(0.89176, 1, 0.53094)
        rgb=(0.89591, 1, 0.52483)
        rgb=(0.9001, 1, 0.51868)
        rgb=(0.90432, 1, 0.51251)
        rgb=(0.90857, 1, 0.5063)
        rgb=(0.91285, 1, 0.50006)
        rgb=(0.91717, 1, 0.49379)
        rgb=(0.92151, 1, 0.48749)
        rgb=(0.92588, 1, 0.48116)
        rgb=(0.93028, 1, 0.4748)
        rgb=(0.93472, 1, 0.46841)
        rgb=(0.93918, 1, 0.46199)
        rgb=(0.94368, 1, 0.45554)
        rgb=(0.9482, 1, 0.44906)
        rgb=(0.95276, 1, 0.44255)
        rgb=(0.95734, 1, 0.436)
        rgb=(0.96196, 1, 0.42943)
        rgb=(0.96661, 1, 0.42282)
        rgb=(0.97129, 1, 0.41619)
        rgb=(0.976, 1, 0.40952)
        rgb=(0.98074, 1, 0.40283)
        rgb=(0.98551, 1, 0.3961)
        rgb=(0.99031, 1, 0.38934)
        rgb=(0.99514, 1, 0.38255)
        rgb=(1, 1, 0.37573)
        rgb=(1, 0.99511, 0.37378)
        rgb=(1, 0.99018, 0.37182)
        rgb=(1, 0.98523, 0.36986)
        rgb=(1, 0.98025, 0.36791)
        rgb=(1, 0.97523, 0.36595)
        rgb=(1, 0.97019, 0.36399)
        rgb=(1, 0.96511, 0.36204)
        rgb=(1, 0.96, 0.36008)
        rgb=(1, 0.95487, 0.35812)
        rgb=(1, 0.9497, 0.35616)
        rgb=(1, 0.9445, 0.35421)
        rgb=(1, 0.93927, 0.35225)
        rgb=(1, 0.93401, 0.35029)
        rgb=(1, 0.92872, 0.34834)
        rgb=(1, 0.9234, 0.34638)
        rgb=(1, 0.91805, 0.34442)
        rgb=(1, 0.91267, 0.34247)
        rgb=(1, 0.90726, 0.34051)
        rgb=(1, 0.90182, 0.33855)
        rgb=(1, 0.89634, 0.33659)
        rgb=(1, 0.89084, 0.33464)
        rgb=(1, 0.8853, 0.33268)
        rgb=(1, 0.87974, 0.33072)
        rgb=(1, 0.87414, 0.32877)
        rgb=(1, 0.86852, 0.32681)
        rgb=(1, 0.86286, 0.32485)
        rgb=(1, 0.85717, 0.3229)
        rgb=(1, 0.85146, 0.32094)
        rgb=(1, 0.84571, 0.31898)
        rgb=(1, 0.83993, 0.31703)
        rgb=(1, 0.83412, 0.31507)
        rgb=(1, 0.82828, 0.31311)
        rgb=(1, 0.82241, 0.31115)
        rgb=(1, 0.81651, 0.3092)
        rgb=(1, 0.81057, 0.30724)
        rgb=(1, 0.80461, 0.30528)
        rgb=(1, 0.79862, 0.30333)
        rgb=(1, 0.79259, 0.30137)
        rgb=(1, 0.78654, 0.29941)
        rgb=(1, 0.78045, 0.29746)
        rgb=(1, 0.77434, 0.2955)
        rgb=(1, 0.76819, 0.29354)
        rgb=(1, 0.76202, 0.29159)
        rgb=(1, 0.75581, 0.28963)
        rgb=(1, 0.74957, 0.28767)
        rgb=(1, 0.7433, 0.28571)
        rgb=(1, 0.737, 0.28376)
        rgb=(1, 0.73068, 0.2818)
        rgb=(1, 0.72432, 0.27984)
        rgb=(1, 0.71792, 0.27789)
        rgb=(1, 0.7115, 0.27593)
        rgb=(1, 0.70505, 0.27397)
        rgb=(1, 0.69857, 0.27202)
        rgb=(1, 0.69206, 0.27006)
        rgb=(1, 0.68551, 0.2681)
        rgb=(1, 0.67894, 0.26614)
        rgb=(1, 0.67233, 0.26419)
        rgb=(1, 0.6657, 0.26223)
        rgb=(1, 0.65903, 0.26027)
        rgb=(1, 0.65234, 0.25832)
        rgb=(1, 0.64561, 0.25636)
        rgb=(1, 0.63885, 0.2544)
        rgb=(1, 0.63206, 0.25245)
        rgb=(1, 0.62524, 0.25049)
        rgb=(1, 0.6184, 0.24853)
        rgb=(1, 0.61152, 0.24658)
        rgb=(1, 0.6046, 0.24462)
        rgb=(1, 0.59766, 0.24266)
        rgb=(1, 0.59069, 0.2407)
        rgb=(1, 0.58369, 0.23875)
        rgb=(1, 0.57666, 0.23679)
        rgb=(1, 0.56959, 0.23483)
        rgb=(1, 0.5625, 0.23288)
        rgb=(1, 0.55538, 0.23092)
        rgb=(1, 0.54822, 0.22896)
        rgb=(1, 0.54103, 0.22701)
        rgb=(1, 0.53382, 0.22505)
        rgb=(1, 0.52657, 0.22309)
        rgb=(1, 0.51929, 0.22114)
        rgb=(1, 0.51199, 0.21918)
        rgb=(1, 0.50465, 0.21722)
        rgb=(1, 0.49728, 0.21526)
        rgb=(1, 0.48988, 0.21331)
        rgb=(1, 0.48245, 0.21135)
        rgb=(1, 0.47499, 0.20939)
        rgb=(1, 0.4675, 0.20744)
        rgb=(1, 0.45997, 0.20548)
        rgb=(1, 0.45242, 0.20352)
        rgb=(1, 0.44484, 0.20157)
        rgb=(1, 0.43722, 0.19961)
        rgb=(1, 0.42958, 0.19765)
        rgb=(1, 0.42191, 0.19569)
        rgb=(1, 0.4142, 0.19374)
        rgb=(1, 0.40646, 0.19178)
        rgb=(1, 0.3987, 0.18982)
        rgb=(1, 0.3909, 0.18787)
        rgb=(1, 0.38307, 0.18591)
        rgb=(1, 0.37521, 0.18395)
        rgb=(1, 0.36733, 0.182)
        rgb=(1, 0.35941, 0.18004)
        rgb=(1, 0.35146, 0.17808)
        rgb=(1, 0.34347, 0.17613)
        rgb=(1, 0.33546, 0.17417)
        rgb=(1, 0.32742, 0.17221)
        rgb=(1, 0.31935, 0.17025)
        rgb=(1, 0.31125, 0.1683)
        rgb=(1, 0.30311, 0.16634)
        rgb=(1, 0.29495, 0.16438)
        rgb=(1, 0.28675, 0.16243)
        rgb=(1, 0.27853, 0.16047)
        rgb=(1, 0.27027, 0.15851)
        rgb=(1, 0.26199, 0.15656)
        rgb=(1, 0.25367, 0.1546)
        rgb=(1, 0.24532, 0.15264)
        rgb=(1, 0.23694, 0.15068)
        rgb=(1, 0.22853, 0.14873)
        rgb=(1, 0.2201, 0.14677)
        rgb=(1, 0.21163, 0.14481)
        rgb=(1, 0.20312, 0.14286)
        rgb=(1, 0.19459, 0.1409)
        rgb=(1, 0.18603, 0.13894)
        rgb=(1, 0.17744, 0.13699)
        rgb=(1, 0.16882, 0.13503)
        rgb=(1, 0.16016, 0.13307)
        rgb=(1, 0.15148, 0.13112)
        rgb=(1, 0.14277, 0.12916)
        rgb=(1, 0.13402, 0.1272)
        rgb=(1, 0.12524, 0.12524)
        rgb=(0.99315, 0.12329, 0.12329)
        rgb=(0.98627, 0.12133, 0.12133)
        rgb=(0.97936, 0.11937, 0.11937)
        rgb=(0.97242, 0.11742, 0.11742)
        rgb=(0.96545, 0.11546, 0.11546)
        rgb=(0.95845, 0.1135, 0.1135)
        rgb=(0.95141, 0.11155, 0.11155)
        rgb=(0.94435, 0.10959, 0.10959)
        rgb=(0.93726, 0.10763, 0.10763)
        rgb=(0.93013, 0.10568, 0.10568)
        rgb=(0.92298, 0.10372, 0.10372)
        rgb=(0.91579, 0.10176, 0.10176)
        rgb=(0.90857, 0.099804, 0.099804)
        rgb=(0.90133, 0.097847, 0.097847)
        rgb=(0.89405, 0.09589, 0.09589)
        rgb=(0.88674, 0.093933, 0.093933)
        rgb=(0.8794, 0.091977, 0.091977)
        rgb=(0.87203, 0.09002, 0.09002)
        rgb=(0.86463, 0.088063, 0.088063)
        rgb=(0.8572, 0.086106, 0.086106)
        rgb=(0.84974, 0.084149, 0.084149)
        rgb=(0.84225, 0.082192, 0.082192)
        rgb=(0.83473, 0.080235, 0.080235)
        rgb=(0.82718, 0.078278, 0.078278)
        rgb=(0.81959, 0.076321, 0.076321)
        rgb=(0.81198, 0.074364, 0.074364)
        rgb=(0.80434, 0.072407, 0.072407)
        rgb=(0.79666, 0.07045, 0.07045)
        rgb=(0.78896, 0.068493, 0.068493)
        rgb=(0.78122, 0.066536, 0.066536)
        rgb=(0.77345, 0.064579, 0.064579)
        rgb=(0.76566, 0.062622, 0.062622)
        rgb=(0.75783, 0.060665, 0.060665)
        rgb=(0.74997, 0.058708, 0.058708)
        rgb=(0.74208, 0.056751, 0.056751)
        rgb=(0.73416, 0.054795, 0.054795)
        rgb=(0.72621, 0.052838, 0.052838)
        rgb=(0.71823, 0.050881, 0.050881)
        rgb=(0.71022, 0.048924, 0.048924)
        rgb=(0.70218, 0.046967, 0.046967)
        rgb=(0.6941, 0.04501, 0.04501)
        rgb=(0.686, 0.043053, 0.043053)
        rgb=(0.67787, 0.041096, 0.041096)
        rgb=(0.6697, 0.039139, 0.039139)
        rgb=(0.66151, 0.037182, 0.037182)
        rgb=(0.65328, 0.035225, 0.035225)
        rgb=(0.64503, 0.033268, 0.033268)
        rgb=(0.63674, 0.031311, 0.031311)
        rgb=(0.62842, 0.029354, 0.029354)
        rgb=(0.62008, 0.027397, 0.027397)
        rgb=(0.6117, 0.02544, 0.02544)
        rgb=(0.60329, 0.023483, 0.023483)
        rgb=(0.59485, 0.021526, 0.021526)
        rgb=(0.58638, 0.019569, 0.019569)
        rgb=(0.57788, 0.017613, 0.017613)
        rgb=(0.56935, 0.015656, 0.015656)
        rgb=(0.56079, 0.013699, 0.013699)
        rgb=(0.5522, 0.011742, 0.011742)
        rgb=(0.54357, 0.0097847, 0.0097847)
        rgb=(0.53492, 0.0078278, 0.0078278)
        rgb=(0.52624, 0.0058708, 0.0058708)
        rgb=(0.51752, 0.0039139, 0.0039139)
        rgb=(0.50878, 0.0019569, 0.0019569)
        rgb=(0.5, 0, 0)
    }
}

\maketitle
\thispagestyle{empty}
\pagestyle{empty}

\begin{abstract}
In future wireless communication networks, existing active localization will gradually evolve into more sophisticated (passive) sensing functionalities.
One main enabler for this process is the merging of information collected from the network's nodes, sensing the environment in a multi-static deployment.
The current literature considers single sensing node systems and/or single target scenarios, mainly focusing on specific issues pertaining to hardware impairments or algorithmic challenges.\newline
In contrast, in this work we propose an ensemble of techniques for processing the information gathered from multiple sensing nodes, jointly observing an environment with multiple targets.
A scattering model is used within a flexibly configurable framework to highlight the challenges and issues with algorithms used in this distributed sensing task.\newline
We validate our approach by supporting it with detailed link budget evaluations, considering practical millimeter-wave systems' capabilities.
Our numerical evaluations are performed in an indoor scenario, sweeping  a variety of parameter to analyze the KPIs sensitivity with respect to each of them.
The proposed algorithms to fuse information by multiple nodes show significant gains in terms of targets' localization performance, with up to 35\% for the probability of detection, compared to the baseline with a mono-static setup.


\end{abstract}


\acresetall

\blfootnote{This work has been submitted to the IEEE for possible publication. Copyright may be transferred without notice, after which this version may no longer be accessible.}

\section{Introduction}\label{sec:introduction}

The next generation of cellular wireless communication systems 6G, like the previous ones, will bring the necessary evolution regarding transfer speed, latency, etc. for the upcoming years~\cite{viswanathan2020communications}.
On top of this, a novel topic called \ac{isac} draws a lot of attention in the research community.
In this vision, the network incorporates sensing capabilities on top of the already existing communications~\cite{TWild2021}.
This is enabled by the improvements in hardware performance, with massive antenna arrays and aggregation of new frequency bands with enormous bandwidths.
One obvious implementation would incorporate radar functionality for the first time in cellular networks, directly in the existing system, making it cheaper and more sustainable.
The ability of the network to localize and classify even passive objects -- compared to 5G's active localization -- would be a major game changer for specialized services, general improvement of the quality of service, and a plethora of other use cases imaginable~\cite{mandelli2023survey,zhang2021enabling}.\newline
Current research in the field of \ac{isac} is concerned with all kinds of related research topics, essential for a later successful deployment.
Examples are improved power allocation between the communications and radar functionalities~\cite{powerallocation} and optimizing angular sampling~\cite{mandelli_sara}.
A more system-centric view is given in \cite{ramos_autoencoder} and \cite{muth_autoencoder} with the detection and angulation of one and multiple targets with the help of a single mono-static system setup and \ac{dl}.
Given the inherent network structure of cellular systems, it seems beneficial to leverage this multi-static setup to improve coverage and reliability of future sensing setups. 
The authors of \cite{favarelli_trackingfusion} demonstrate the tracking of a single target observed by multiple mono-static systems via different tracking algorithms.
In~\cite{yajnanarayana2023multistatic}, a single transmitter illuminates the scene while multiple receivers extract and fuse the observation with the help of \ac{dl}.
\newline
From the mentioned literature, we deduce that a holistic view on the multi-static sensing problem is still missing.
We propose a model-based end-to-end methodology to process and fuse sensing acquisitions from multi-static operations, locating multiple targets in the environment.
To the best of our knowledge, this is the first work in \ac{isac} literature targeting multi-static sensing for multiple target localization.\newline
This work contributes by highlighting localization limitations, challenges, solutions, and by presenting:
\begin{itemize}
    \item a complete (classical) sensing pipeline, resembling existing cellular networks,
    \item an ensemble of techniques to perform multi-node peak extraction and fusion,
    \item numerical evaluations in a realistic multiple scattering targets scenario with focus on physical practicability and consistency at millimeter wave frequencies.
\end{itemize}
Our results show the gains and benefits from single-node to multi-node fusion, and performance for different constraints regarding link budget and system parameters.\newline
The paper structure is as follows:
Section~\ref{sec:model} introduces the system model and existing \ac{ofdm} radar processing.
On top of this, Section~\ref{sec:framework} describes the methodology necessary for the complete multi-static sensing task.
Section~\ref{sec:scenario} provides information on the investigated scenario, the modelled effects, and its limitations.
After presenting our results in Section~\ref{sec:results}, a summary concludes the paper (Section~\ref{sec:conclusion}).

\begin{figure}[h]
    \centering
    \def\svgwidth{\columnwidth}
\begingroup%
  \makeatletter%
  \providecommand\color[2][]{%
    \errmessage{(Inkscape) Color is used for the text in Inkscape, but the package 'color.sty' is not loaded}%
    \renewcommand\color[2][]{}%
  }%
  \providecommand\transparent[1]{%
    \errmessage{(Inkscape) Transparency is used (non-zero) for the text in Inkscape, but the package 'transparent.sty' is not loaded}%
    \renewcommand\transparent[1]{}%
  }%
  \providecommand\rotatebox[2]{#2}%
  \newcommand*\fsize{\dimexpr\f@size pt\relax}%
  \newcommand*\lineheight[1]{\fontsize{\fsize}{#1\fsize}\selectfont}%
  \ifx\svgwidth\undefined%
    \setlength{\unitlength}{362.05530842bp}%
    \ifx\svgscale\undefined%
      \relax%
    \else%
      \setlength{\unitlength}{\unitlength * \real{\svgscale}}%
    \fi%
  \else%
    \setlength{\unitlength}{\svgwidth}%
  \fi%
  \global\let\svgwidth\undefined%
  \global\let\svgscale\undefined%
  \makeatother%
  \begin{picture}(1,0.82985608)%
    \lineheight{1}%
    \setlength\tabcolsep{0pt}%
    \put(0,0){\includegraphics[width=\unitlength,page=1]{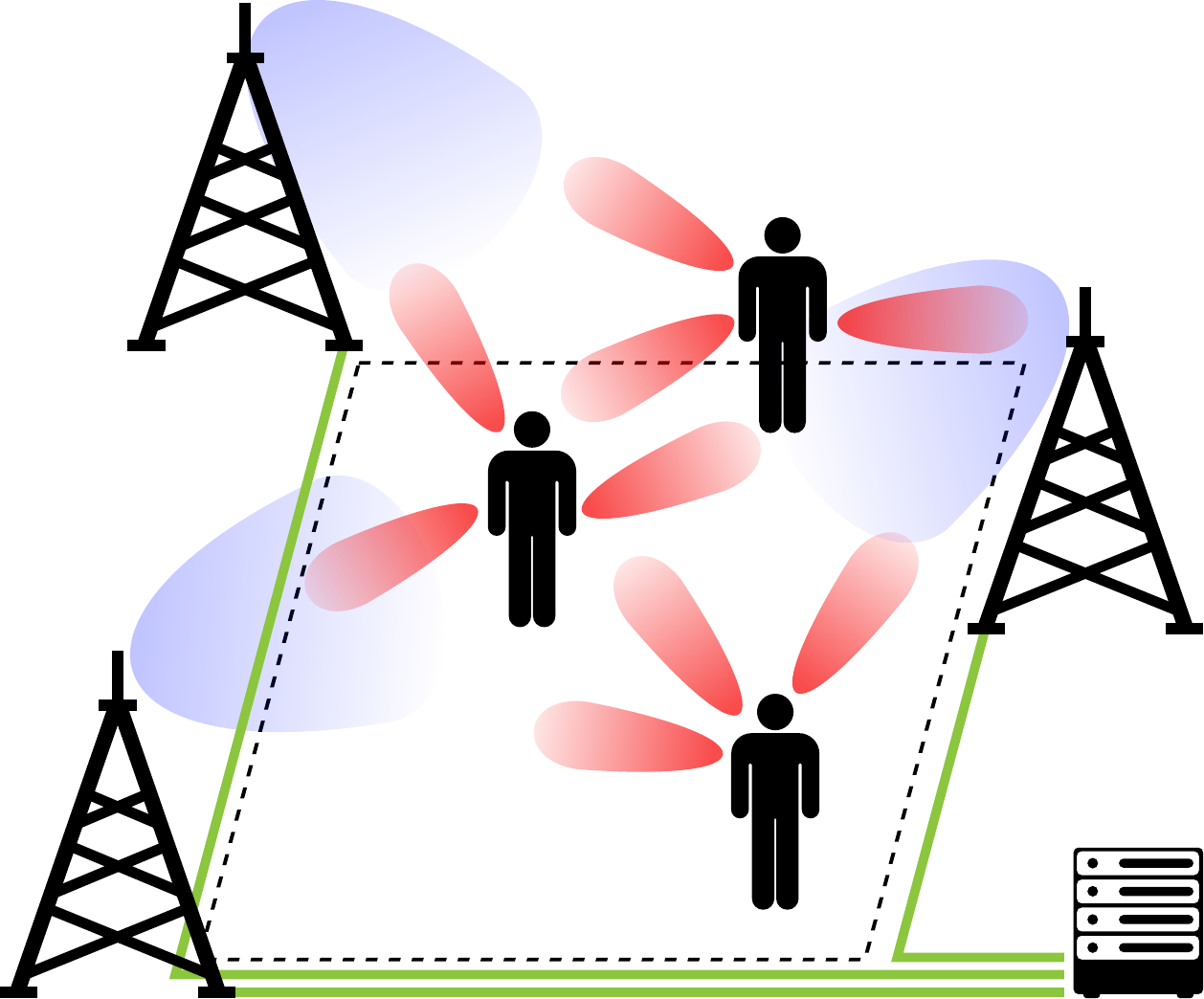}}%
    \put(0.08188367,0.26064238){\color[rgb]{0,0,0}\makebox(0,0)[rt]{\lineheight{1.25}\smash{\begin{tabular}[t]{r}$\text{SAP}_\text{3}$\end{tabular}}}}%
    \put(0.65039226,0.43745453){\color[rgb]{0,0,0}\makebox(0,0)[t]{\lineheight{1.25}\smash{\begin{tabular}[t]{c}$\text{Target}_{1}$\end{tabular}}}}%
    \put(0.44223031,0.28156258){\color[rgb]{0,0,0}\makebox(0,0)[t]{\lineheight{1.25}\smash{\begin{tabular}[t]{c}$\text{Target}_{2}$\end{tabular}}}}%
    \put(0.64437118,0.04528109){\color[rgb]{0,0,0}\makebox(0,0)[t]{\lineheight{1.25}\smash{\begin{tabular}[t]{c}$\text{Target}_{3}$\end{tabular}}}}%
    \put(0.94633704,0.13307245){\color[rgb]{0,0,0}\makebox(0,0)[t]{\lineheight{1.25}\smash{\begin{tabular}[t]{c}SeMF\end{tabular}}}}%
    \put(0.91929087,0.56225212){\color[rgb]{0,0,0}\makebox(0,0)[lt]{\lineheight{1.25}\smash{\begin{tabular}[t]{l}$\text{SAP}_\text{1}$\end{tabular}}}}%
    \put(0.18661742,0.79851336){\color[rgb]{0,0,0}\makebox(0,0)[rt]{\lineheight{1.25}\smash{\begin{tabular}[t]{r}$\text{SAP}_\text{2}$\end{tabular}}}}%
  \end{picture}%
\endgroup%

    \caption{Proposed sensing network structure with targets and nodes/\acrfullpl{sap} connected to the central \acrfull{semf}. The transmitted wavefront is indicated in blue, the back-scattered received signal in red.}
    \label{fig:blockdigram_model}
\end{figure}
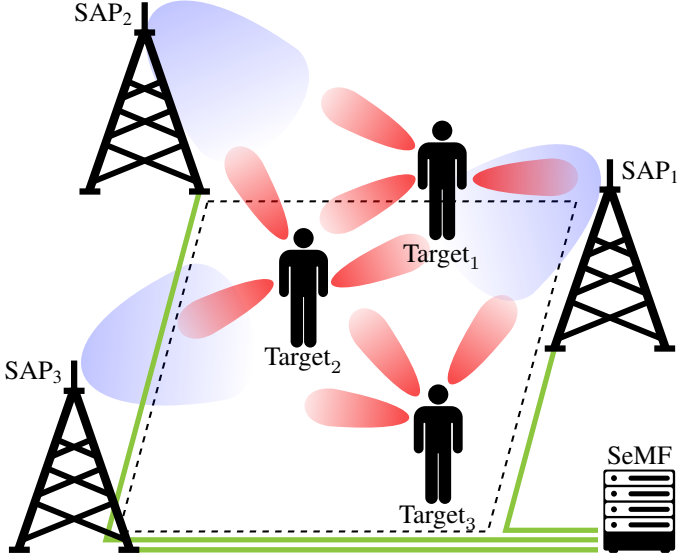

\section{\acrshort{ofdm} Radar and System Model}\label{sec:model}
In this work we study wireless network based radar detection and localization using multiple spatially distributed nodes, referred to as \acp{sap}, that could represent any access point or user terminal performing sensing operations.
We consider \ac{ofdm} radar~\cite{ofdm_radar}, compliant with cellular network waveforms~\cite{3gpp_38211}, for the parameter estimation within each \ac{sap}.\newline
While we focus on extracting single snapshot (azimuth) angle and range information for localization, an extension to include elevation angle and Doppler information would be straightforward.

\subsection{\acrshort{ofdm} Radar}
For each \ac{sap}, we consider a typical \ac{ofdm} system operating at a carrier frequency $f_c$ with bandwidth $B$ and $N_\text{sub}$ subcarriers, spaced by $\Delta f = \nicefrac{B}{N_\text{sub}}$.
Each \ac{sap} senses the environment with mono-static acquisitions, thus just receiving the signal transmitted at the same site.
The \ac{tx} is a single illuminating antenna, while the \ac{rx} \ac{ula} comprises $K$ antenna elements with spacing $d$.
Note that the achievable sensing resolution depends just on the sum co-array between transmitter and receiver~\cite{hoctor1990unifying}, thus it is equivalent for any combination of \ac{tx} and \ac{rx} with a total number of elements equal to $k + 1$.\newline
The transmitted complex symbols modulated onto each subcarrier $n$ are denoted with $\vec{x} = \left[ x_1, x_2, \dots, x_N \right]$ and the received signal is $\vec{y} = \left[ y_1, y_2, \dots, y_N \right]$.
We define the $N_\text{sub} \times K$ \ac{ctf} matrix $\mathbf{H}$. The entry corresponding to the $n$-th subcarrier and $k$-th antenna is deduced from the frequency domain response of the superposition of signal reflections from each scatterer $p \in \mathcal{P}$ in the environment
\begin{equation}
    h_{n,k} = \sum_{p \in \mathcal{P}} \alpha_p e^{j 2 \pi \left(- \frac{2n\Delta f}{c} l_p + \frac{d k f_c}{c} \sin \left( \theta_p \right) \right )}
    \label{eq:ctf}
\end{equation}
The complex coefficient $\alpha_p$ of each path is influenced by a variety of parameters, further described in Sec.~\ref{subsec:budget}.
The phase shift over the subcarriers is determined by the path's propagation distance $l_p$, corresponding to twice the target distance to the \ac{sap}, while the phase shift over the antennas is determined by the impinging angle $\theta_p$ of the path on the \ac{ula}.
The vector $\vec{z} = \left[ z_1, z_2, \dots, z_N \right]$ represents random complex \ac{awgn} samples drawn from ${z_n \sim \mathcal{C}\mathcal{N}(0,\,\sigma^{2})}$, resulting in a noise power over the whole bandwidth of $P_N = \sigma^{2} N_\text{sub}$.\newline
Accordingly, the noisy received \ac{ofdm} symbols are given by
\begin{equation}
    y_{n,k} = x_n \cdot h_{n,k} + z_{n,k} \; .
    \label{eq:system_model}
\end{equation}
Single-tap equalization is applied to extract an estimate of the \ac{ctf} $\mathbf{\hat{H}}$, assuming known transmit symbols, through
\begin{equation}
    \hat{h}_{n,k} = \frac{ y_{n,k}}{x_n} \; .
    \label{eq:equalization}
\end{equation}

\subsection{Periodogram}
We estimate the angle and range of targets from the \ac{ctf} by periodogram computation, as described in~\cite{ofdm_radar}.
Angular information is inferred from 
the \ac{dft}, where the linear phase shifts over the receiving \ac{ula} elements due to the azimuth angle of arrival.
Range information is derived via the inverse \ac{dft} from the phase shift over consecutive subcarriers~\cite{braun_diss}.
The two \ac{dft}-based operations can be combined for joint angle-range estimation through
\begin{equation}
    \mathrm{P}_{n',k'} = \frac{1}{N'K'} \left| \sum_{m=0}^{N'-1}\left(\sum_{i=0}^{K'-1} \hat{h}_{m,i} e^{-\mathrm{j}2\pi\frac{ik'}{K'} }\right) e^{\mathrm{j}2\pi\frac{mn'}{N'} } \right|^ 2 \; .
\end{equation}
The \ac{ctf} is zero-padded to shape $N' \times K'$ for simple global interpolation.
Suitable \ac{dft} parametrization ensures mapping from respective bins to the correct angle and range labels.

\subsection{Signal-to-Noise Ratio}
We define the symbol-wise communication-centric \ac{snr} $\gamma_\mathrm{com}$ with the received symbol power $P_{S,\mathrm{com}}$ as
\begin{equation}
    \gamma_\mathrm{com}
    = \frac{ P_{S,\mathrm{com}}}{P_N}
\end{equation}
For target extraction, the respective peak with power $P_{S,\mathrm{imag}}$ in the periodogram is to be considered.
With the processing gain provided by the \acp{dft} size in the periodogram~\cite{braun_diss}, this raises the \ac{snr} to the sensing \ac{snr} given by
\begin{equation}
    \gamma_\mathrm{imag}
    = \frac{P_{S,\mathrm{imag}}}{P_N}
    = \frac{P_{S,\mathrm{com}}}{P_N} \cdot N_\mathrm{sub} K \; .
\end{equation}
For the case of a single impulsive target with no (i.e., rectangular) windowing applied, $\gamma_\mathrm{imag}$ corresponds to the ratio between periodogram peak power and noise floor power.
Otherwise, the peak power and processing gain is decreased.


\section{Network Structure and Processing}\label{sec:framework}

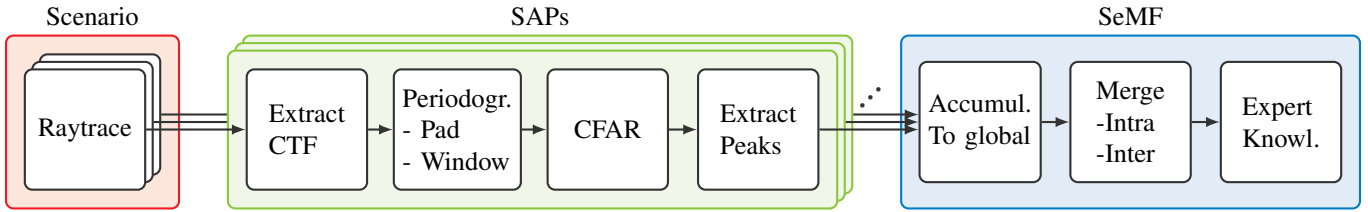
\begin{figure*}[t]
	\centering
  	\begin{tikzpicture}


    \def\height{1.6cm}
    \def\width{1.6cm}
    
    \pgfdeclarelayer{layer0}
    \pgfdeclarelayer{layer1}
    \pgfdeclarelayer{layer2}
    \pgfdeclarelayer{layer3}
    \pgfdeclarelayer{layer4}
    \pgfsetlayers{main,layer0,layer1,layer2,layer3,layer4}
    
    \begin{pgfonlayer}{layer0}
        \node[draw=uni_rot,fill=uni_rot!10, thick, rounded corners=.1cm, minimum height=\height+0.5cm+0.2cm,minimum width=\width+0.2cm+0.5cm,align=left] (scen) at (1.15,1.15) {};

        \node[draw=uni_apfelgruen,fill=uni_apfelgruen!10, thick, rounded corners=.1cm, minimum height=\height+0.5cm,minimum width=0.5cm+4*\width+3*0.4cm,align=left] (sap1) at (7+0.2,1.25) {};
        \node[draw=uni_apfelgruen,fill=uni_apfelgruen!10, thick, rounded corners=.1cm, minimum height=\height+0.5cm,minimum width=0.5cm+4*\width+3*0.4cm,align=left] (sap2) at (7+0.1,1.15) {};
        \node[draw=uni_apfelgruen,fill=uni_apfelgruen!10, thick, rounded corners=.1cm, minimum height=\height+0.5cm,minimum width=0.5cm+4*\width+3*0.4cm,align=left] (sap3) at (7,1.05) {};
        
        \node[draw=uni_mittelblau,fill=uni_mittelblau!10, thick, rounded corners=.1cm, minimum height=\height+0.5cm+0.2cm,minimum width=0.5cm+3*\width+2*0.4cm,align=left] (semf) at (14.95+0,1.15) {};
    \end{pgfonlayer}
    
    \begin{pgfonlayer}{layer1}
        \node[draw=unigrau,fill=white, thick, rounded corners=.1cm, minimum height=\height,minimum width=\width,align=left] (scenb1) at (1.25,1.25) {};
        \node[draw=unigrau,fill=white, thick, rounded corners=.1cm, minimum height=\height,minimum width=\width,align=left] (scenb2) at (1.15,1.15) {};
        \node[draw=unigrau,fill=white, thick, rounded corners=.1cm, minimum height=\height,minimum width=\width,align=left] (scenb3) at (1.05,1.05) {Raytrace};
    \end{pgfonlayer}
    
    \begin{pgfonlayer}{layer2}
        \node[draw=unigrau,fill=white, thick, rounded corners=.1cm, minimum height=\height,minimum width=\width,align=left] (sapb1) at (4,1.05) {Extract\\\acrshort{ctf}};
        \node[draw=unigrau,fill=white, thick, rounded corners=.1cm, minimum height=\height,minimum width=\width,align=left] (sapb2) at (6,1.05) {Periodogr.\\- Pad\\- Window};
        \node[draw=unigrau,fill=white, thick, rounded corners=.1cm, minimum height=\height,minimum width=\width,align=left] (sapb3) at (8,1.05) {\acrshort{cfar}};
        \node[draw=unigrau,fill=white, thick, rounded corners=.1cm, minimum height=\height,minimum width=\width,align=left] (sapb4) at (10,1.05) {Extract\\Peaks};
    \end{pgfonlayer}
        
    \begin{pgfonlayer}{layer3}
        \node[draw=unigrau,fill=white, thick, rounded corners=.1cm, minimum height=\height,minimum width=\width,align=left] (semfb1) at (12.95,1.15) {Accumul.\\To global}; 
        \node[draw=unigrau,fill=white, thick, rounded corners=.1cm, minimum height=\height,minimum width=\width,align=left] (semfb2) at (14.95,1.15) {Merge\\-Intra\\-Inter};
        \node[draw=unigrau,fill=white, thick, rounded corners=.1cm, minimum height=\height,minimum width=\width,align=left] (semfb3) at (16.95,1.15) {Expert\\Knowl.};
    \end{pgfonlayer}
    
    \begin{pgfonlayer}{layer4}
        \draw[thick, unigrau] (scenb1.east) -- ($(sap3.west) + (0,0.2)$);
        \draw[thick, unigrau] (scenb2.east) -- ($(sap3.west) + (0,0.1)$);
        \draw[-latex,thick, unigrau] (scenb3.east) -- ($(sapb1.west) + (0,0)$);
        
        \draw[-latex,thick, unigrau] (sapb1.east) -- (sapb2.west);
        \draw[-latex,thick, unigrau] (sapb2.east) -- (sapb3.west);
        \draw[-latex,thick, unigrau] (sapb3.east) -- (sapb4.west);
        \draw[thick, unigrau] (sapb4.east) -- (sap3.east);
        
        \node[rotate=45, unigrau] at ($(sap1.east) + (0.25,0.25)$) {\textbf{\dots}};
        \draw[-latex,thick, unigrau] (sap1.east) -- ($(semfb1.west) + (0,0.1)$);
        \draw[-latex,thick, unigrau] (sap2.east) -- ($(semfb1.west) + (0,0)$);
        \draw[-latex,thick, unigrau] (sap3.east) -- ($(semfb1.west) + (0,-0.1)$);
        
        \draw[-latex,thick, unigrau] (semfb1.east) -- (semfb2.west);
        \draw[-latex,thick, unigrau] (semfb2.east) -- (semfb3.west);
        
        \node[anchor=south] at (scen.north) {Scenario};
        \node[anchor=south] at ($(sap1.north) + (-0.1,0)$) {\acrshortpl{sap}};
        \node[anchor=south] at (semf.north) {\acrshort{semf}};
    \end{pgfonlayer}

\end{tikzpicture}
	\caption{Processing pipeline from raytracing of scenario to acquisitions in \acrfullpl{sap} and overall estimate in the \acrfull{semf}.}
	\label{fig:processing_blockdiagram}
\end{figure*}

Our implemented network structure resembles already existing cellular networks with multiple \acp{sap} and a central \ac{semf} located in the core network, similarly to the 5G \ac{lmf} for active localization introduce by 3GPP with Release~15~\cite{3gpp_rel15}.
Fig.~\ref{fig:processing_blockdiagram} depicts the flow of information within the scenario and the channel model, the target extraction process inside the individual \acp{sap}, and the aggregation in the \ac{semf}.
Here, compared to the mono- and bi-static case with a single pair of \ac{tx} and \ac{rx}, we want to highlight the challenges and proposed solutions which enable the multi-static sensing task.

\subsection{Scenario}
This interchangeable block provides the received signal for each \ac{sap} based on the scenario.
For the contribution of this paper we found sufficient to leverage a simplified raytracer to determine the different channels, similar to related literature~\cite{muth_autoencoder, powerallocation, favarelli_trackingfusion, yajnanarayana2023multistatic}. 
In principle, any channel model, fully fledged raytracers~\cite{arnold_maxray} or even measured data could be injected at this point.\\
The basis is a fully configurable scenario, with targets and arbitrary numbers and configurations of \ac{tx} and \ac{rx} antenna arrays.
The simple raytracer determines all paths between antenna elements and the $\lvert \mathcal{P} \rvert$ scattering points, and imposes propagation specific effects as described in Sec.~\ref{sec:scenario}.
Based on these paths, for one colocated \ac{tx}-\ac{rx} antenna array combination corresponding to a \ac{sap}, the static and bandwidth-limited \ac{ctf} $\mathbf{H}$ is calculated.
Finally, we multiply the \ac{ctf} with the transmitted \ac{ofdm} symbol and add noise, according to Eq.~\eqref{eq:system_model}.

\subsection{Sensing Access Point (\acrshort{sap})}
We assume a bandwidth-limited link between the \acp{sap} and the \ac{semf}, which is to be expected in reality.
Given the system parameters, raw IQ samples would easily exceed 100\,Gbit/s (8\,antennas\,\texttimes\,800\,MHz bandwidth\,\texttimes\,10\,bit resolution\,\texttimes\,2\,IQ) and periodogram-based signalling would be beyond 50\,Mbit/image (128\,beams\,\texttimes\,2984\,subcarriers\,\texttimes\,64\,bit resolution\,\texttimes\,2\,IQ).
This renders full processing in the \ac{semf} infeasible and requires extraction to be done in the individual \acp{sap}.
Given the colocated \ac{tx} and \ac{rx} in each \ac{sap} and therefore known transmit signal, the noisy bandwidth limited \ac{ctf} $\vec{\hat{h}}$ can be extracted from the received signal -- provided by the scenario -- via Eq.~\eqref{eq:equalization}.
To facilitate later interpolation in amplitude and location, each symbol, i.e., dimension of antennas and subcarriers, is padded by at least a factor of four.
Windowing with a Chebyshev window ensures control over constant sidelobe-levels.
We fixed the sidelobe attenuation to 30\,dB for low resolution degradation while still ensuring that targets in reasonable distance are not covered by sidelobes.
We use a statistical \ac{cfar} threshold approach for detection and extraction of peaks~\cite{richards_principles}.
The power threshold $\zeta_\mathrm{\ac{cfar}}$ is determined with a certain probability of false alarm $P_{FA}$ of detecting targets in a noisy but target-less scenario with
\begin{equation}
    \zeta_\mathrm{\acrshort{cfar}} = \sqrt{-P_N \ln{\left( 1 - \left( 1 - P_{FA} \right) ^\frac{1}{N_\text{sub} K} \right)}} \; .
\end{equation}
We assume that the noise power is known by taking the ground truth added $P_N$, which is consistent with the noise power in the periodogram.
In reality, this has to be estimated, for example from target less bins or by the communications part of the system.
Further comparing this threshold to the expected sidelobe level of the maximum peak $\zeta_{SL,\mathrm{max}}$, including an empirically found factor $\kappa=4$, decreases the probability of detecting sidelobes and their constructive inter-target interference in high \ac{snr} regimes
\begin{equation}
    \zeta'_\mathrm{\acrshort{cfar}} = \max{ \{ \zeta_\mathrm{\ac{cfar}}, \kappa \cdot \zeta_{SL,\mathrm{max}} \} } \; .
\end{equation}
While there are other extraction approaches such as \mbox{(OS-)}\ac{cfar}~\cite{richards_radar}, they are heavily dependent on suitable parameterization for the given scenario, which is an open research topic and not within the scope of this work.\\
For the extraction itself, we chose binary successive cancellation~\cite{braun_diss}. 
We extract the maximum bin of the periodogram and cancel it by setting an ellipsoidal region based on the peak size to zero.
This is iteratively repeated for all peaks until no bin exceeds the previously set $\zeta'_\mathrm{\acrshort{cfar}}$.
We found that typical coherent cancellation algorithms designed for impulsive targets~\cite{braun_diss} do not yield the desired suppression effect. This is due to the fact that targets might be distributed in larger areas than the resolution limit of the system, generating a distributed -- and not impulsive -- echo.
The extracted peak positions and powers are quadratically interpolated to reduce scalloping loss.
Both are sent to the \ac{semf} together with information such as noise power and peak size, amongst others.\\
Given our focus on consistent power considerations throughout the processing, up to the periodogram noise floor, we can infer realistic system performance under the to be expected noise power induced by the receiver.

\subsection{Sensing Management Function (\acrshort{semf})}
The \ac{semf} accumulates all extracted targets from the different \acp{sap} to be subsequently replaced by estimates in the fusion process.
Each target is transformed to global coordinates according to its parent \ac{sap} location and rotation.
Several steps are applied to account for the non-perfect extraction process:
First, we apply a geometric intra-\ac{sap} clustering via DBSCAN~\cite{db_scan}.
We take double the range resolution as the merging distance. 
Given the applied windowing, this fuses those extracted targets closer together than physically resolvable.
The subsequent inter-\ac{sap} fusion process relies on the same algorithm for geometric merging between peaks of different \acp{sap}.\newline
To reduce false positives, we neglect peaks outside the investigated room size or ``multi-node" requirement, filtering out peaks not observed by at least two \acp{sap}.\newline
The main idea is to use more and more information available from the extracted peaks to do more intelligent fusion in the future, e.g., by incorporating the peak power and shape.
Statistical approaches could include metrics already existing in standards such as angle/range resolution, in addition to new metrics, like the probability that a certain target truly exists.

\section{Scenario Setup}\label{sec:scenario}
\begin{figure*}[t]
	\centering
  	\input{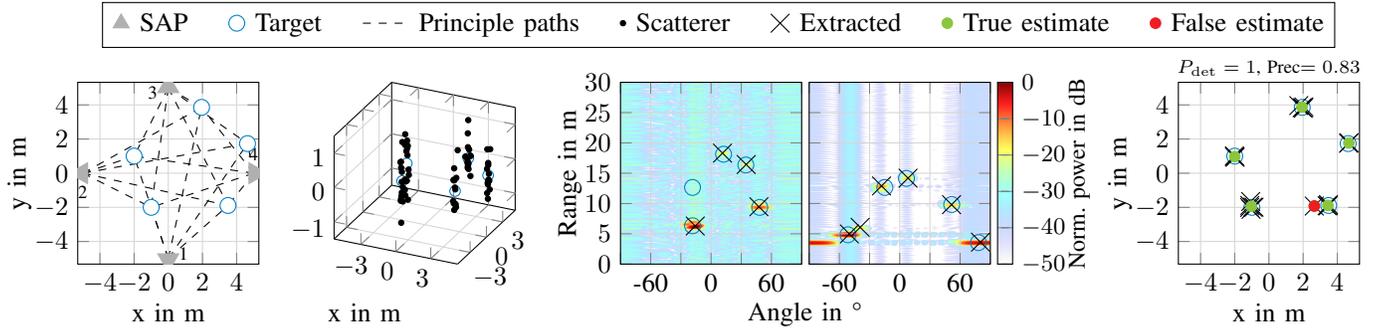}
	\caption{Example plots from scenario (left to right): (1) \acrfullpl{sap} with principle paths to targets (top view, for visualization), (2) traced scatterers around targets, (3) \ac{sap}\textsubscript{1} periodogram with ground truth, extracted targets, and -56\,dBm noise power, (4) \ac{sap}\textsubscript{4} periodogram with -76\,dB noise power, (5) resulting scenario top view after merging of four \acp{sap}.}
	\label{fig:2_scenario_examples}
\end{figure*}

We conduct our investigation in a factory-like free-space scenario as proposed in the majority of current literature, e.g.,~\cite{muth_autoencoder, powerallocation, favarelli_trackingfusion, yajnanarayana2023multistatic}.
The scene is observed by up to four \acp{sap} positioned centered on each wall. 
Each consists of a single room-illuminating \ac{tx} and a $\lambda/2$-spaced \ac{rx} \ac{ula}, both with patch radiation patterns for the individual elements.\newline
Up to eight targets are placed in random positions inside a 10\,\texttimes\,10\,\texttimes\,3\,m cube with no minimal target separation distance.
To mimic radar fluctuations, the targets consist of 15 individual scattering points which together resemble the \ac{rcs} of a human with 1\,m\textsuperscript{2}.
The scattering points are Gaussian distributed in x-, y-, and z-direction with standard deviations of 0.1, 0.03, and 0.5\,m, respectively.\newline
The simulation calculates each path with the assumption of free-space propagation (path-loss exponent $\eta=2$) including the typical losses and gains for such a system~\cite{mandelli2023survey}. 
Occlusions from other targets are accounted for by intersection with their scatter point bounding box.\newline
Fig.~\ref{fig:2_scenario_examples} shows the scenario with exemplary targets, observed periodograms, as well as extracted and fused results.
The periodograms show different occuring effects, such as non-ellipsoidal peaks from the scattering model, occlusion in the first and multiple extractions in the second, leading to a decreased precision in the evaluation.

\subsection{Limitations}\label{subsec:lim}
This static scenario to investigate fundamental sensing performance contains a number of limitations.
The simple raytracer model assumes free-space propagation without any clutter, no near-field effects, and no multipath effects between similar constant \ac{rcs} targets justified with the high losses at mmWave frequencies.
The \acp{sap} are assumed to be full-duplex capable, i.e., there is no self-interference due to the colocated \ac{tx} and \ac{rx}.\newline
The transmitted \ac{ofdm} signal is modulated with a constant power scheme such as 4-QAM and is assumed orthogonal between \acp{sap} via time- or frequency-division multiplexing.
The single-snapshot merging of two-dimensional extractions is purely geometry-based with suitable, empirically found merging distances.\newline
Nonetheless, we believe that the mentioned points do not diminish the validity of this work, and that it can serve as a baseline for future research in the area of multi-node sensing.

\subsection{Link Budget}\label{subsec:budget}
\begin{figure}[h]
	\centering
	\begin{tikzpicture}


\def\axisoffset{0.1cm}
\def\linkwidth{1.0cm}
\def\maxy{8cm}
\def\maxx{8.5cm}

\pgfdeclarelayer{background}
\pgfdeclarelayer{ylabels}
\pgfdeclarelayer{axis}
\pgfdeclarelayer{plateaus}
\pgfdeclarelayer{links}
\pgfsetlayers{background,ylabels,axis,plateaus,links,main}

\begin{pgfonlayer}{axis}
    \node[anchor=south, align=center] at (1cm,\maxy) (node_yaxis) {power\\in dBm};
    \draw[-latex, ultra thick] (1cm,-0.8cm) -- (node_yaxis);
\end{pgfonlayer}

\begin{pgfonlayer}{background}
    \node[rectangle, text=black, fill=uni_mittelblau!70, minimum width=\maxx-1cm, minimum height=0.8cm] (r) at (4.75cm,-0.4cm) {thermal noise floor};

    \draw[draw=none, text=black, fill=uni_apfelgruen!10] (1cm,0) rectangle ++(2*\linkwidth-2*\axisoffset,0.96*\maxy) node[midway, yshift=4cm, rotate=0] {\acrshort{tx}};
    \draw[draw=none, text=black, fill=uni_rot!10] (1cm+2*\linkwidth-2*\axisoffset,0) rectangle ++(1*\linkwidth+4*\axisoffset,0.96*\maxy) node[midway, yshift=4cm, rotate=0] {Channel};
    \draw[draw=none, text=black, fill=uni_mittelblau!10] (1cm+3*\linkwidth+2*\axisoffset,0) rectangle ++(\maxx-1cm-3*\linkwidth-2*\axisoffset,0.96*\maxy) node[midway, yshift=4cm, rotate=0] {\acrshort{rx}};
    
    \draw[draw=none, text=black, fill=uni_gelb!50] (1cm+5.5*\linkwidth-0.3cm,0.05*\maxy) rectangle ++(0.6cm,0.15*\maxy) node[midway, rotate=90] {$\gamma_\mathrm{com}$};
    \draw[draw=none, text=black, fill=uni_gelb!50] (1cm+6.5*\linkwidth-0.3cm,0.05*\maxy) rectangle ++(0.6cm,0.33*\maxy) node[midway, rotate=90] {$\gamma_\mathrm{imag}$};
\end{pgfonlayer}

\begin{pgfonlayer}{ylabels}
    \node[anchor=east] at (1cm-\axisoffset,0.89*\maxy) {$P_\text{\acrshort{tx}}$};
    \node[anchor=east] at (1cm-\axisoffset,0.65*\maxy) {$0$};
    \node[anchor=east] at (1cm-\axisoffset,0.2*\maxy) {$P_\text{\acrshort{rx}}$};
    \node[anchor=east] at (1cm-\axisoffset,0.05*\maxy) {$P_N$};
    \node[anchor=east] at (1cm-\axisoffset,0) {$P_n$};
\end{pgfonlayer}

\begin{pgfonlayer}{plateaus}
    \draw[dashed,black] (1cm-\axisoffset,0.65*\maxy) -- (1cm+2*\linkwidth+\axisoffset,0.65*\maxy);
    \draw[dashed, black] (1cm-\axisoffset,0.2*\maxy) -- (1cm+5*\linkwidth-\axisoffset,0.2*\maxy);
    \draw[black] (1cm-\axisoffset,0.05*\maxy) -- (\maxx,0.05*\maxy);
    \draw[black] (1cm-\axisoffset,0) -- (\maxx,0cm);
    
    \draw[thick, uni_rot] (1cm-\axisoffset,0.89*\maxy) -- +(\axisoffset+\linkwidth+\axisoffset,0cm);
    \draw[thick, uni_rot] (1cm+\linkwidth-\axisoffset,0.95*\maxy) -- +(\axisoffset+\linkwidth+\axisoffset,0cm);
    \draw[thick, uni_rot] (1cm+2*\linkwidth-\axisoffset,0.5*\maxy) -- +(\axisoffset+\linkwidth+\axisoffset,0cm);
    \draw[thick, uni_rot] (1cm+3*\linkwidth-\axisoffset,0.55*\maxy) -- +(\axisoffset+\linkwidth+\axisoffset,0cm);
    \draw[thick, uni_rot] (1cm+4*\linkwidth-\axisoffset,0.1*\maxy) -- +(\axisoffset+\linkwidth+\axisoffset,0cm);
    \draw[thick, uni_rot] (1cm+5*\linkwidth-\axisoffset,0.2*\maxy) -- +(\axisoffset+\linkwidth+\axisoffset,0cm);
    \draw[thick, uni_rot] (1cm+6*\linkwidth-\axisoffset,0.38*\maxy) -- +(\axisoffset+\linkwidth+\axisoffset,0cm);
\end{pgfonlayer}

\begin{pgfonlayer}{links}
    \draw[-latex, thick] (1cm+\linkwidth,0.89*\maxy) -- (1cm+\linkwidth,0.95*\maxy) node[midway, right, rotate=0] {$G_\text{\acrshort{tx}}$};
    \draw[-latex, thick] (1cm+2*\linkwidth,0.95*\maxy) -- (1cm+2*\linkwidth,0.5*\maxy) node[midway, right, rotate=0] {$G_P$};
    \draw[-latex, thick] (1cm+3*\linkwidth,0.5*\maxy) -- (1cm+3*\linkwidth,0.55*\maxy) node[midway, right, rotate=0] {$G_\text{\acrshort{rcs}}$};
    \draw[-latex, thick] (1cm+4*\linkwidth,0.55*\maxy) -- (1cm+4*\linkwidth,0.1*\maxy) node[midway, right, rotate=0] {$G_P$};
    \draw[-latex, thick] (1cm+5*\linkwidth,0.1*\maxy) -- (1cm+5*\linkwidth,0.2*\maxy) node[midway, left, rotate=0] {$G_\text{\acrshort{rx}}$};
    \draw[-latex, thick] (1cm+6*\linkwidth,0.2*\maxy) -- (1cm+6*\linkwidth,0.38*\maxy) node[midway, left, rotate=0] {$G_\text{Per}$};
    \draw[-latex, thick] (1cm+5.5*\linkwidth,0*\maxy) -- (1cm+5.5*\linkwidth,0.05*\maxy) node[midway, right, rotate=0] {$F$};

\end{pgfonlayer}

\end{tikzpicture} 
	\caption{Schematic of link budget model}
	\label{fig:linkbudget_model}
\end{figure}
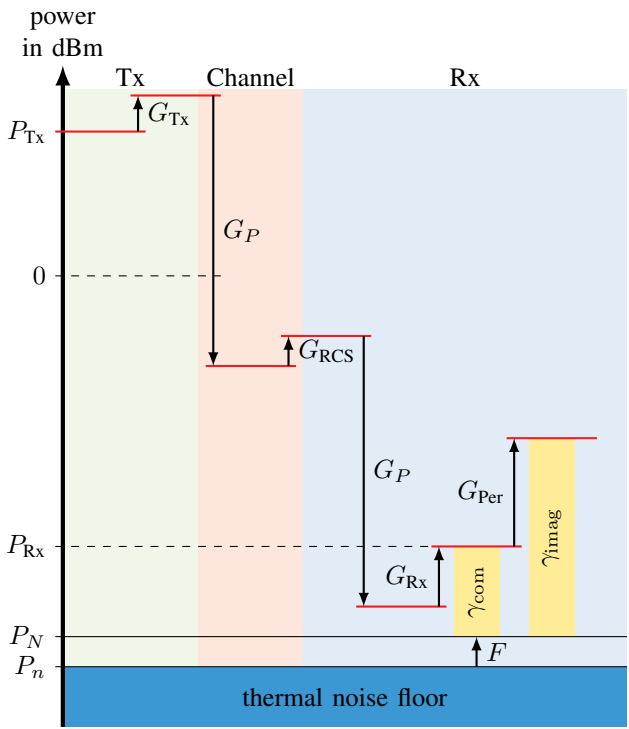

In this subsection, we discuss results without sweeping noise power, adopting values that should resemble the performance of a real world system as close as possible.
Therefore, we introduce the link budget model from Fig.~\ref{fig:linkbudget_model}.
It depicts the flow of power -- with losses and gains -- along the propagation path and also serves for validation purposes.
The individual parameters with descriptions and values for the link budget considerations and simulations are given in Tab.~\ref{tab:linkbudget_parameters}.
\begin{table}[h]
	\caption{Parameters for Link Budget Model and Simulations}
	\label{tab:linkbudget_parameters}
	
\begin{tabular}{p{0.08\columnwidth}p{0.57\columnwidth}p{0.2\columnwidth}}
    \toprule
    Variable                      & Description                               & Value         \\ \midrule
    $P_\text{\acrshort{tx}}$      & \acrshort{tx} power for whole array       & 29\,dBm       \\
    $G_\text{\acrshort{tx}}$      & \Acrshort{tx} antenna gain (here: single patch) & 6.6\,dBi \\
    $G_P$                         & Path gain (one way)                       & \textit{variable}  \\
    $G_\text{\acrshort{rcs}}$     & \Acrlong{rcs}                             & 1\,m\textsuperscript{2}  \\
    $G_\text{\acrshort{rx}}$      & \Acrshort{rx} antenna gain (here: 16\,\texttimes\,1 patch array)  & 18.6\,dBi      \\    
    $P_\text{\acrshort{rx}}$      & \Acrshort{rx} power                       & \textit{variable} \\
    $G_\text{Per}$                & Processing gain of periodogram            & 42\,dB          \\
    $\gamma_\mathrm{com}$         & \Acrlong{snr} (communication signal)      & \textit{variable}  \\
    $\gamma_\mathrm{imag}$        & \Acrlong{snr} (periodogram)               & $\gamma_\mathrm{com} + G_\text{Per}$ \\
    $B$                           & System bandwidth                          & 800\,MHz                \\
    $F$                           & \Acrlong{rx} \acrlong{nf}                 & 8\,dB         \\
    $P_n$                         & Thermal noise power for $B$ and 21\,\textdegree C & -84.9\,dBm  \\
    $P_N$                         & Equivalent noise power                    & $P_n+F=-76.9\,\text{dBm}$ \\ \bottomrule
\end{tabular}

\end{table}

\subsection{Evaluation metrics}
For this passive localization task, we use three evaluation metrics.
Let $\vec{o}$ denote one (estimated) target location vector, one element from a set $\mathcal{O}$ with cardinality $\lvert \mathcal{O} \rvert$.\newline
With the probability of detection $P_\mathrm{det}$, we evaluate the correct detections, i.e., the true positives $\vec{o}_{\mathrm{true}}^+ \in \mathcal{O}_{\mathrm{true}}^+$ with respect to the ground truth actual positives $\vec{o}^+ \in \mathcal{O}^+$.
For a detected element $\vec{o}_{\mathrm{det}}^+ \in \mathcal{O}_{\mathrm{det}}^+$ to be part of the true positives $\vec{o}_{\mathrm{det}}^+ \in \mathcal{O}_{\mathrm{true}}^+ \subseteq \mathcal{O}_{\mathrm{det}}^+$, it must not be farther located than distance $r$ away from the next undetected ground truth target
\begin{equation}
    \left\| \vec{o}_{\mathrm{det}}^+ - \vec{o}^+ \right\|_2 \le r \; ,
\end{equation}
with distance given by the $\ell^2$-norm $\| \cdot \|_2$.
Thus, we get
\begin{equation}
    P_\mathrm{det} = \frac{| \mathcal{O}_{\mathrm{true}}^+ |}{| \mathcal{O}^+ |} \; .
\end{equation}
We cannot rasterize the scenario in e.g., range bins to infer the set of actual negatives $\mathcal{O}^-$ due to the multiple overlapping \acp{sap}.
Therefore, to assess false detections, we use the precision $\rho$ which relates the true positives to the overall detected targets
\begin{equation}
    \rho = \frac{| \mathcal{O}_{\mathrm{true}}^+ |}{| \mathcal{O}_{\mathrm{det}}^+ |} \; .
\end{equation}
The $F_1$-score combines the two metrics through their harmonic mean with
\begin{equation}
    F_1 = \frac{2 P_\mathrm{det} \rho}{P_\mathrm{det} + \rho}
\end{equation}
and serves as a combined performance metric.

\section{Results}\label{sec:results}
In the following, we summarize our findings for different configurations and their performance trade-offs with respect to multi-\ac{sap} fusion, number of antenna elements, bandwidth, target density, and room size.\newline
In contrast to the \ac{snr} sweep typically done in communications, we vary the noise power at the \ac{rx} array of each \ac{sap}.
This is due to the fact that the \ac{snr} changes based on the received signal power, which is, among other things, dependent on the targets' scattering characteristics and their distances to the \acp{sap}.
If no noise power value is explicitly given, we determine the operating point through the link budget with thermal noise floor according to the given bandwidth and noise figure (e.g. 800\,MHz: -84.9\,dBm + 8\,dB).


\subsection{Multi-\acrshort{sap} Fusion Performance}
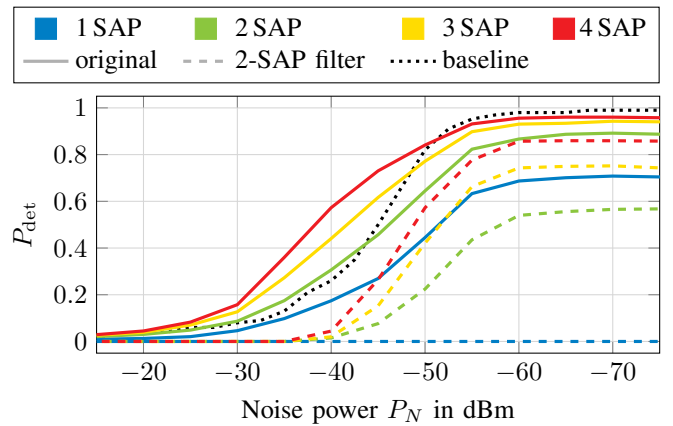
\begin{figure}[h]
	\centering
  	\begin{tikzpicture}
	\begin{axis}[
		width=0.5\textwidth, height=5cm,
		grid=major, grid style={solid,unigrau!20},
		xmin=-75,xmax=-15,
		ymin=-0.05,ymax=1.05,
		xtick={-90,-80,...,-10},
		x dir=reverse,
		ytick={0,0.2,...,1},
		xlabel={Noise power $P_N$ in dBm}, ylabel={$P_\mathrm{det}$},
		xlabel near ticks, ylabel near ticks,
		legend cell align={left},
		legend style={/tikz/every even column/.append style={column sep=2.5mm}},
	 	legend columns=4,
		legend style={at={(1,1.05)},anchor=south east},
		]

        \addlegendimage{only marks, mark=square*, uni_mittelblau, mark size=4pt}
        \addlegendentry{1\,\ac{sap}}
        \addlegendimage{only marks, mark=square*, uni_apfelgruen, mark size=4pt}
        \addlegendentry{2\,\ac{sap}}
        \addlegendimage{only marks, mark=square*, uni_gelb, mark size=4pt}
        \addlegendentry{3\,\ac{sap}}
        \addlegendimage{only marks, mark=square*, uni_rot, mark size=4pt}
        \addlegendentry{4\,\ac{sap}}
        
        \addlegendimage{darkgray176, solid, very thick}
        \addlegendentry{original}
        \addlegendimage{darkgray176, dashed, very thick}
        \addlegendentry{2-\ac{sap} filter}
        \addlegendimage{black, dotted, very thick}
        \addlegendentry{baseline}

        \addplot[black, dotted, very thick] table[x=noisepower, y=p_det_1, col sep=comma]{./tikz/data/sweep_noisepower_1sap_100rep_bound.dat}; 
        
        \addplot[uni_mittelblau, very thick] table[x=noisepower, y=p_det_1, col sep=comma]{./tikz/data/sweep_noisepower_4sap_1merge_100rep.dat}; 
		\addplot[uni_apfelgruen, very thick] table[x=noisepower, y=p_det_2, col sep=comma]{./tikz/data/sweep_noisepower_4sap_1merge_100rep.dat}; 
        \addplot[uni_gelb, very thick] table[x=noisepower, y=p_det_3, col sep=comma]{./tikz/data/sweep_noisepower_4sap_1merge_100rep.dat}; 
        \addplot[uni_rot, very thick] table[x=noisepower, y=p_det_4, col sep=comma]{./tikz/data/sweep_noisepower_4sap_1merge_100rep.dat}; 
		
		\addplot[uni_mittelblau, dashed, very thick] table[x=noisepower, y=p_det_1, col sep=comma]{./tikz/data/sweep_noisepower_4sap_2merge_100rep.dat}; 
        \addplot[uni_apfelgruen, dashed, very thick] table[x=noisepower, y=p_det_2, col sep=comma]{./tikz/data/sweep_noisepower_4sap_2merge_100rep.dat}; 
        \addplot[uni_gelb, dashed, very thick] table[x=noisepower, y=p_det_3, col sep=comma]{./tikz/data/sweep_noisepower_4sap_2merge_100rep.dat}; 
        \addplot[uni_rot, dashed, very thick] table[x=noisepower, y=p_det_4, col sep=comma]{./tikz/data/sweep_noisepower_4sap_2merge_100rep.dat}; 
        
	\end{axis}
\end{tikzpicture}
	\caption{Probability of detection for varying noise power and number of \acrfullpl{sap}.}
	\label{fig:fusion_pdet}
\end{figure}
In general, Fig.~\ref{fig:fusion_pdet} shows higher probabilities of detection for lower noise powers.
The 70\% for the single \ac{sap} case is consistent with related work~\cite{muth_autoencoder}. 
The implemented fusion process of multiple \acp{sap} yields a probability of detection gain throughout the whole noise power range.
This improvement thanks to fusion is most significant with 27\% from a single to two \acp{sap} and reduces to 2\% from three to four nodes in the saturated low noise power regime.\newline
The dotted baseline curve is supposed to approximate a theoretically best achievable performance for a single \ac{sap} given a single, non scattering target, estimated by taking the maximum peak in the periodogram.
While we do not beat this very simplified reference throughout the whole range with multiple \acp{sap}, we observe gains from -20 to -50\,dBm and come close for lower noise powers. This additionally gained reliability/certainty can be leveraged in different ways, with the most obvious one being a reduction in the number of false detections.
\begin{figure}[h]
	\centering
  	\begin{tikzpicture}
	\begin{axis}[
		width=0.5\textwidth, height=5cm,
		grid=major, grid style={solid,unigrau!20},
		xmin=-78,xmax=-15,
		ymin=-0.05,ymax=1.05,
		x dir=reverse,
		xtick={-90,-80,...,-10},
        ytick={0,0.2,...,1},
		xlabel={Noise power $P_N$ in dBm},
		ylabel={Precision},
		xlabel near ticks, ylabel near ticks,
		legend cell align={left},
		legend style={/tikz/every even column/.append style={column sep=2.5mm}},
	 	legend columns=4,
		legend style={at={(1,1.05)},anchor=south east},
		]

        \addlegendimage{only marks, mark=square*, uni_mittelblau, mark size=4pt}
        \addlegendentry{1\,\ac{sap}}
        \addlegendimage{only marks, mark=square*, uni_apfelgruen, mark size=4pt}
        \addlegendentry{2\,\ac{sap}}
        \addlegendimage{only marks, mark=square*, uni_gelb, mark size=4pt}
        \addlegendentry{3\,\ac{sap}}
        \addlegendimage{only marks, mark=square*, uni_rot, mark size=4pt}
        \addlegendentry{4\,\ac{sap}}
        
        \addlegendimage{darkgray176, solid, very thick}
        \addlegendentry{original}
        \addlegendimage{darkgray176, dashed, very thick}
        \addlegendentry{2-\ac{sap} filter}
        \addlegendimage{black, dotted, very thick}
        \addlegendentry{baseline}

        \addplot[black, dotted, very thick] table[x=noisepower, y=p_prec_1, col sep=comma]{./tikz/data/sweep_noisepower_1sap_100rep_bound.dat}; 
        
        \addplot[uni_mittelblau, very thick] table[x=noisepower, y=p_prec_1, col sep=comma]{./tikz/data/sweep_noisepower_4sap_1merge_100rep.dat}; 
        \addplot[uni_apfelgruen, very thick] table[x=noisepower, y=p_prec_2, col sep=comma]{./tikz/data/sweep_noisepower_4sap_1merge_100rep.dat}; 
		\addplot[uni_gelb, very thick] table[x=noisepower, y=p_prec_3, col sep=comma]{./tikz/data/sweep_noisepower_4sap_1merge_100rep.dat}; 
		\addplot[uni_rot, very thick] table[x=noisepower, y=p_prec_4, col sep=comma]{./tikz/data/sweep_noisepower_4sap_1merge_100rep.dat}; 

		\addplot[uni_mittelblau, dashed, very thick] table[x=noisepower, y=p_prec_1, col sep=comma]{./tikz/data/sweep_noisepower_4sap_2merge_100rep.dat}; 
        \addplot[uni_apfelgruen, dashed, very thick] table[x=noisepower, y=p_prec_2, col sep=comma]{./tikz/data/sweep_noisepower_4sap_2merge_100rep.dat}; 
        \addplot[uni_gelb, dashed, very thick] table[x=noisepower, y=p_prec_3, col sep=comma]{./tikz/data/sweep_noisepower_4sap_2merge_100rep.dat}; 
        \addplot[uni_rot, dashed, very thick] table[x=noisepower, y=p_prec_4, col sep=comma]{./tikz/data/sweep_noisepower_4sap_2merge_100rep.dat}; 

        
	\end{axis}
\end{tikzpicture}
	\caption{Precision for varying noise power and number of \acrfullpl{sap}.}
	\label{fig:fusion_prec}
\end{figure}
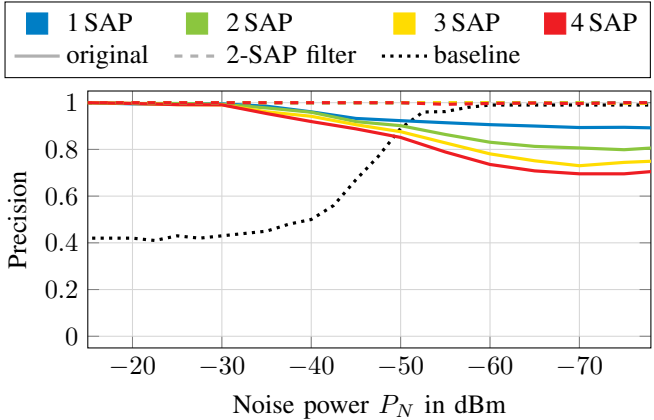
The worse precision performance in the low noise regime in Fig.~\ref{fig:fusion_prec} can be completely traced back to the used scattering model.
Its fluctuation loss and associated non-ellipsoidal peak shape in combination with ellipsoidal binary extraction aggravates correct peak detection.
This reduces precision -- even more for multiple \acp{sap} -- by 1) causing multiple detections in cases where only a single target is present and 2) shifted peak locations outside the target matching radius.
While the 2-\acrshort{sap} filter can mitigate this effect to some extent, we did not give further attention to the open research question of peak extraction.
The baseline curve shows the expected behavior with no false detections for high \acp{snr}.
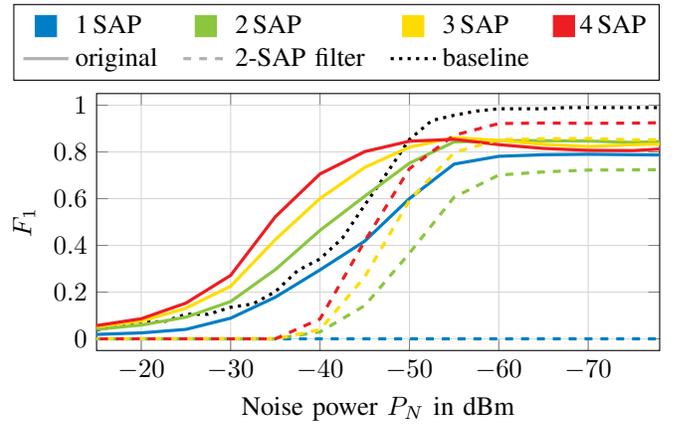
\begin{figure}[h]
	\centering
  	\begin{tikzpicture}
	\begin{axis}[
		width=0.5\textwidth, height=5cm,
		grid=major, grid style={solid,unigrau!20},
		xmin=-78,xmax=-15,
		ymin=-0.05,ymax=1.05,
		xtick={-90,-80,...,-10},
		x dir=reverse,
		ytick={0,0.2,...,1},
		xlabel={Noise power $P_N$ in dBm}, ylabel={$F_1$},
		xlabel near ticks, ylabel near ticks,
		legend cell align={left},
        legend style={/tikz/every even column/.append style={column sep=2.5mm}},
	 	legend columns=4,
		legend style={at={(1,1.05)},anchor=south east},
		]

        \addlegendimage{only marks, mark=square*, uni_mittelblau, mark size=4pt}
        \addlegendentry{1\,\ac{sap}}
        \addlegendimage{only marks, mark=square*, uni_apfelgruen, mark size=4pt}
        \addlegendentry{2\,\ac{sap}}
        \addlegendimage{only marks, mark=square*, uni_gelb, mark size=4pt}
        \addlegendentry{3\,\ac{sap}}
        \addlegendimage{only marks, mark=square*, uni_rot, mark size=4pt}
        \addlegendentry{4\,\ac{sap}}
        
        \addlegendimage{darkgray176, solid, very thick}
        \addlegendentry{original}
        \addlegendimage{darkgray176, dashed, very thick}
        \addlegendentry{2-\ac{sap} filter}
        \addlegendimage{black, dotted, very thick}
        \addlegendentry{baseline}
        
        \addplot[black, dotted, very thick] table[x=noisepower, y=f1_1, col sep=comma]{./tikz/data/sweep_noisepower_1sap_100rep_bound.dat}; 
        
        \addplot[uni_mittelblau, very thick] table[x=noisepower, y=f1_1, col sep=comma]{./tikz/data/sweep_noisepower_4sap_1merge_100rep.dat}; 
		\addplot[uni_apfelgruen, very thick] table[x=noisepower, y=f1_2, col sep=comma]{./tikz/data/sweep_noisepower_4sap_1merge_100rep.dat}; 
		\addplot[uni_gelb, very thick] table[x=noisepower, y=f1_3, col sep=comma]{./tikz/data/sweep_noisepower_4sap_1merge_100rep.dat}; 
        \addplot[uni_rot, very thick] table[x=noisepower, y=f1_4, col sep=comma]{./tikz/data/sweep_noisepower_4sap_1merge_100rep.dat}; 
		
		\addplot[uni_mittelblau, dashed, very thick] table[x=noisepower, y=f1_1, col sep=comma]{./tikz/data/sweep_noisepower_4sap_2merge_100rep.dat}; 
        \addplot[uni_apfelgruen, dashed, very thick] table[x=noisepower, y=f1_2, col sep=comma]{./tikz/data/sweep_noisepower_4sap_2merge_100rep.dat}; 
        \addplot[uni_gelb, dashed, very thick] table[x=noisepower, y=f1_3, col sep=comma]{./tikz/data/sweep_noisepower_4sap_2merge_100rep.dat}; 
        \addplot[uni_rot, dashed, very thick] table[x=noisepower, y=f1_4, col sep=comma]{./tikz/data/sweep_noisepower_4sap_2merge_100rep.dat}; 
        
	\end{axis}
\end{tikzpicture}
	\caption{$F_1$-score for varying noise power and number of \acrfullpl{sap}.}
	\label{fig:fusion_f1}
\end{figure}
Combining both previous metrics to the $F_1$-score in Fig.~\ref{fig:fusion_f1} shows marginal improvements due to the precision issue, which again can be completely mitigated through the 2-\acrshort{sap} filter for a probability of detection converging to 92\% with 4 \acp{sap}.

\subsection{Influence of Bandwidth and Number of Antennas}
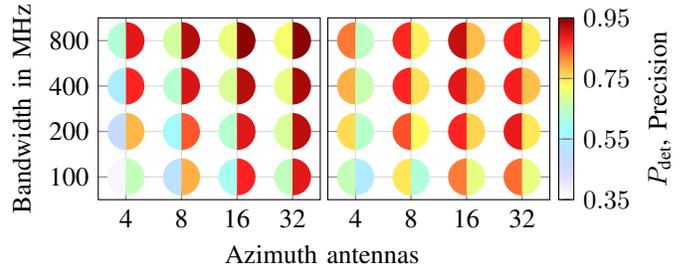
\begin{figure}[h]
	\centering
  	\begin{tikzpicture}

    \def\mincbar{0.35}
    \def\maxcbar{0.95}

    \pgfdeclareplotmark{halfcircle}{%
        \pgfusepathqfillstroke
        \pgfsetfillcolor{mapped color}%
        \pgfpathmoveto{\pgfpoint{0}{\pgfplotmarksize}}
        \pgfpatharc{90}{270}{\pgfplotmarksize}
        \pgfpathclose
        \pgfusepathqfill
    }

    \centering
    \begin{groupplot}[
        group style={group size=2 by 1, vertical sep=0mm, horizontal sep=1mm},
        width=0.25\textwidth, height=4cm,
        grid=major, grid style={solid,unigrau!20},
        xmin=-0.5, xmax=3.5, xlabel near ticks,
        xtick={0,1,2,3},
        xticklabels={4,8,16,32},
        xlabel near ticks,
        ymin=-0.5, ymax=3.5, ylabel near ticks,
        ytick={0,1,2,3},
        yticklabels={100,200,400,800},
        ylabel near ticks,
        colormap name=jet_inue,
		colorbar style={ylabel={$P_\text{det}$, Precision}, at={(1.04,0)}, anchor=south west, width=2mm, ytick={0.35,0.55,...,0.95}},
		legend style={fill=white, fill opacity=0.4, draw opacity=1, text opacity=1, nodes={scale=0.6, transform shape}, at={(1,1.03)}, anchor=south, /tikz/every even column/.append style={column sep=0.1cm}},
		legend cell align={left},
		legend columns=4,
		view={0}{90}, ]
    
        \nextgroupplot[ylabel={Bandwidth in MHz}, ylabel style = {yshift=-0mm}, xlabel={Azimuth antennas}, xlabel style = {xshift=15mm}, point meta min=\mincbar, point meta max=\maxcbar, ]
            
            \addplot[scatter, mark=halfcircle, only marks, mark size=7pt, point meta=explicit] table[x=ant_ind, y=bw_ind, meta=p_det_1, col sep=comma] {tikz/data/sweep_antbw_4sap_100rep.dat};

            \addplot[scatter, mark=halfcircle, mark options={rotate=180}, only marks, mark size=7pt, point meta=explicit] table[x=ant_ind, y=bw_ind, meta=p_det_3, col sep=comma] {tikz/data/sweep_antbw_4sap_100rep.dat};

        \nextgroupplot[yticklabels=\empty, colorbar, point meta min=\mincbar, point meta max=\maxcbar,]
            
            \addplot[scatter, mark=halfcircle, only marks, mark size=7pt, point meta=explicit] table[x=ant_ind, y=bw_ind, meta=p_prec_1, col sep=comma] {tikz/data/sweep_antbw_4sap_100rep.dat};

            \addplot[scatter, mark=halfcircle, mark options={rotate=180}, only marks, mark size=7pt, point meta=explicit] table[x=ant_ind, y=bw_ind, meta=p_prec_3, col sep=comma] {tikz/data/sweep_antbw_4sap_100rep.dat};
            
	\end{groupplot}
\end{tikzpicture}
	\caption{Probability of detection (left) and Precision (right) for different configurations of number of antenna elements and bandwidth. Left half-circle with single node, right half-circle with three node fusion. Noise power is determined by the bandwidth.}
	\label{fig:bw_ant}
\end{figure}

Fig.~\ref{fig:bw_ant} depicts the operating points for different antenna and bandwidth configurations in the \ac{sap}.
The link budget parameters for antenna gain and thermal noise power are adapted accordingly.
More antennas and more bandwidth yield a higher probability of detection by providing higher resolutions in their respective domains.
This allows resolving closely spaced targets and increases peak isolation in the periodogram.
Inherently, the peak power is also positively influenced through the higher processing gain, resulting in better extraction performance overall.\newline
The smallest configuration with 4 antennas and 100\,MHz bandwidth results in 37\% detection probability.
Increasing the antenna size or bandwidth 8-fold yields 65\% and 62\%, respectively, and their combination up to 72\% in probability of detection.
Again, leveraging the fusion of multiple \acp{sap} provides an additional performance increase of 25 -- 35\%.\newline
While precision also improves up to 78\% with additional antennas and higher bandwidth, at a certain point performance decreases again.
This is due to the resolution being so high that the individual scattering points have significant impact on the peak shape.
Then, simple peak extraction, even more so by ellipsoidal binary cancellation, is not sufficient anymore.
Better metrics than precision (and also detection with respect to location) -- which is suitable only for impulsive scattering objects or low enough resolution -- have to be found.

\subsection{Dependence on Number of Targets}
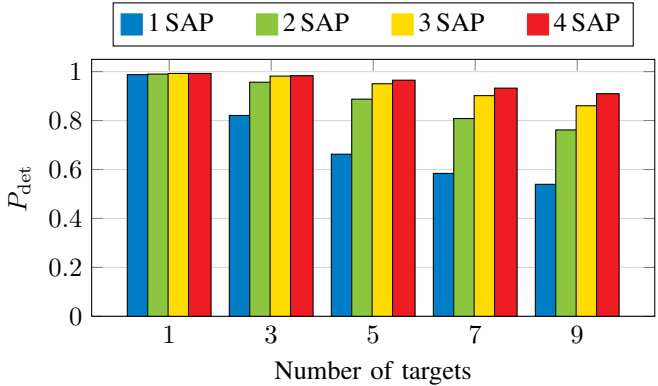
\begin{figure}[h]
	\centering
  	\begin{tikzpicture}
	\begin{axis}[
		width=0.5\textwidth, height=5cm,
		ymajorgrids, grid style={solid,unigrau!20},
		ymin=-0,ymax=1.05,
		xtick={1,3,5,7,9},
		ytick={0,0.2,...,1},
		xlabel={Number of targets}, ylabel={$P_\mathrm{det}$},
		xlabel near ticks, ylabel near ticks,
		bar width=3mm,
		legend cell align={left},
		legend style={/tikz/every even column/.append style={column sep=0.5cm}},
	 	legend columns=4,
		legend style={at={(0.5,1.05)},anchor=south},
		]

        \addlegendimage{only marks, mark=square*, mark size=4pt, color=uni_mittelblau}
        \addlegendentry{1\,\acrshort{sap}}
        \addlegendimage{only marks, mark=square*, mark size=4pt, color=uni_apfelgruen}
        \addlegendentry{2\,\acrshort{sap}}
        \addlegendimage{only marks, mark=square*, mark size=4pt, color=uni_gelb}
        \addlegendentry{3\,\acrshort{sap}}
        \addlegendimage{only marks, mark=square*, mark size=4pt, color=uni_rot}
        \addlegendentry{4\,\acrshort{sap}}
        
        \addplot[ybar,fill=uni_mittelblau] coordinates {
            (0.4,0.9875)  %
            (2.4,0.8208333333333333)
            (4.4,0.6625)
            (6.4,0.5839285714285714)
            (8.4,0.5394444444444444)
            };
            
        \addplot[ybar,fill=uni_apfelgruen] coordinates {
            (0.8,0.99)
            (2.8,0.9566666666666666)
            (4.8,0.8875)
            (6.8,0.8082142857142857)
            (8.8,0.7616666666666667)
            };
        
        \addplot[ybar,fill=uni_gelb] coordinates {
            (1.2,0.9925)
            (3.2,0.9816666666666666)  %
            (5.2,0.9505)
            (7.2,0.9017857142857143)
            (9.2,0.8605555555555557)
            };
        
        \addplot[ybar,fill=uni_rot] coordinates {
            (1.6,0.9925)
            (3.6,0.9833333333333333)
            (5.6,0.9650000000000002)
            (7.6,0.9325)
            (9.6,0.9097222222222223)
            };
        
        
	\end{axis}
\end{tikzpicture}
	\caption{Probability of detection for varying number of targets and number of \acrfullpl{sap}.}
	\label{fig:nb_tar}
\end{figure}
Running the simulations for a fixed number of targets in Fig.~\ref{fig:nb_tar} shows an overall decrease in probability of detection with increasing number of targets and decreasing number of \acp{sap}.
This can be attributed to a higher likelihood of targets being occluded by others, as well as a higher probability of being closer together than resolvable by the system.\newline
Again, the benefits of multi-static sensing become clear, as the fusion of multiple \acp{sap} with their different observation points can mitigate this effect to some extent.
For instance, by increasing the \ac{sap} number from one to four, the 44\% loss in performance for going from 1 to 9 targets can be reduced to only 8\%.
\subsection{Influence of Room Size}
\begin{figure}[h]
	\centering
  	\begin{tikzpicture}
	\begin{axis}[
		width=0.5\textwidth, height=5cm,
		grid=major, grid style={solid,unigrau!20},
		xmin=2,xmax=100,
		ymin=-0.05,ymax=1.05,
		xtick={10,20,...,100},
		ytick={0,0.2,...,1},
		xlabel={Room size in m}, ylabel={Probability}, 
		xlabel near ticks, ylabel near ticks,
		legend cell align={left},
		legend style={/tikz/every even column/.append style={column sep=2mm}},
	 	legend columns=5,
		legend style={at={(1,1.05)},anchor=south east},
		]

        \addlegendimage{only marks, mark=square*, uni_mittelblau, mark size=4pt}
        \addlegendentry{1\,\ac{sap}}
        \addlegendimage{only marks, mark=square*, uni_rot, mark size=4pt}
        \addlegendentry{4\,\ac{sap}}
        
        \addlegendimage{darkgray176, solid, very thick}
        \addlegendentry{$P_\mathrm{det}$}
        \addlegendimage{darkgray176, dashed, very thick}
        \addlegendentry{Precision}
        \addlegendimage{black, dotted, very thick}
        \addlegendentry{$P_\mathrm{occ}$}
        
        \addplot[uni_mittelblau, very thick] table[x=roomsize, y=p_det_1, col sep=comma]{./tikz/data/sweep_roomsize_4sap_100rep.dat}; 
        \addplot[uni_rot, very thick] table[x=roomsize, y=p_det_4, col sep=comma]{./tikz/data/sweep_roomsize_4sap_100rep.dat}; 

        \addplot[uni_mittelblau, dashed, very thick] table[x=roomsize, y=p_prec_1, col sep=comma]{./tikz/data/sweep_roomsize_4sap_100rep.dat}; 
        \addplot[uni_rot, dashed, very thick] table[x=roomsize, y=p_prec_4, col sep=comma]{./tikz/data/sweep_roomsize_4sap_100rep.dat}; 

        \addplot[black, dotted, very thick] table[x=roomsize, y=p_occl, col sep=comma]{./tikz/data/sweep_roomsize_4sap_100rep.dat}; 
        
	\end{axis}
\end{tikzpicture}
	\caption{Performance for varying room sizes and number of \acrfullpl{sap}.}
	\label{fig:room_size}
\end{figure}
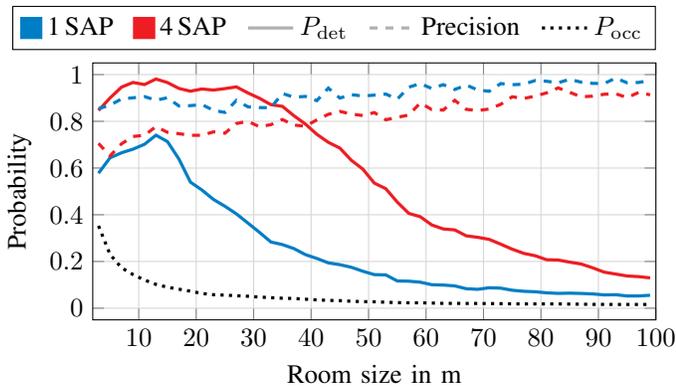
Finally, Fig.~\ref{fig:room_size} shows the performance for different (quadratic) room side lengths.
With increasing room size, the percentage of occluded targets $P_\mathrm{occ}$ decreases due to the bigger area in which the 1 to 8 random targets can be placed.
For a single \ac{sap}, the best probability of detection is achieved around 13\,m room side-length and drops rapidly afterwards.
Here, the probability of occluded and therefore undetectable targets is relatively small, and they are still close enough to the \ac{sap} to allow correct detection.
Multiple \acp{sap} on the other hand can maintain their higher detection probability for a larger range of room sizes from 10 -- 35\,m with a slower decrease afterwards.
This is reasoned with the higher probability that targets are located in a good detection position for at least one \ac{sap}, which further underlines the advantages of fusing multiple \acp{sap}. \newline
The dynamic range of detecting far targets in presence of a close strong target is limited by the window sidelobe attenuation.
To facilitate higher detectable ranges, transmitter beamforming with a \ac{ula} in combination with beam steering to scan the room would be necessary.

\section{Conclusion}\label{sec:conclusion}
In this work, we defined a framework with all the necessary steps to perform sensing of multiple targets in a multi-static scenario, from the creation of a non-impulsive scatterer scenario, over peak extraction and fusion.\newline
It was shown that fusing diverse acquisitions from multiple \acp{sap} increases the overall system performance and robustness significantly. 
Our approach achieves gains of up to 35\% in probability of detection, demonstrating the practical advantages over a single node setup, which is more prone to e.g., occluded or weakly reflecting targets.\newline
Future investigations will focus on optimizing the different subsystems, especially the extraction process, and on more sophisticated fusion techniques. 
We also plan to expand the setup to more realistic raytraced and measured scenarios, including clutter, as well as the fusion of bi-static acquisitions.

\section*{Acknowledgments}
This work was developed within the KOMSENS-6G project, partly funded by the German Ministry of Education and Research under grant 16KISK112K.

\bibliography{references}
\bibliographystyle{IEEEtran}

\end{document}